\documentclass{aastex63}

\usepackage{amsmath}
\usepackage{mathtools}

\usepackage{xcolor}
\newcommand{\COMON}{\begin{color}{blue}}
\newcommand{\COMOFF}{\end{color}}

\usepackage{rotating}

\received{October 13, 2021}
\revised{November 19, 2021}
\accepted{November 24, 2021}
\submitjournal{ApJ}

\shorttitle{ICME Magnetic Complexity Changes Observed by Radially-Aligned Spacecraft}
\shortauthors{Scolini et al.}

\begin{document}

\title{Causes and Consequences of Magnetic Complexity Changes within Interplanetary \\
Coronal Mass Ejections: a Statistical Study}

\correspondingauthor{Camilla Scolini}
\email{camilla.scolini@unh.edu}

\author[0000-0002-5681-0526]{Camilla Scolini}
\affiliation{Institute for the Study of Earth, Oceans, and Space, University of New Hampshire, Durham, NH, USA}
\affiliation{CPAESS, University Corporation for Atmospheric Research, Boulder, CO, USA}

\author[0000-0002-9276-9487]{R\'{e}ka M. Winslow}
\affiliation{Institute for the Study of Earth, Oceans, and Space, University of New Hampshire, Durham, NH, USA}

\author[0000-0002-1890-6156]{No\'{e} Lugaz}
\affiliation{Institute for the Study of Earth, Oceans, and Space, University of New Hampshire, Durham, NH, USA}

\author[0000-0001-6813-5671]{Tarik M. Salman}
\affiliation{Institute for the Study of Earth, Oceans, and Space, University of New Hampshire, Durham, NH, USA}

\author[0000-0001-9992-8471]{Emma E. Davies}
\affiliation{Institute for the Study of Earth, Oceans, and Space, University of New Hampshire, Durham, NH, USA}

\author[0000-0003-3752-5700]{Antoinette B. Galvin}
\affiliation{Institute for the Study of Earth, Oceans, and Space, University of New Hampshire, Durham, NH, USA}



\begin{abstract}
We present the first statistical analysis of complexity changes affecting the magnetic structure of interplanetary coronal mass ejections (ICMEs), with the aim of answering the questions:
How frequently do ICMEs undergo magnetic complexity changes during propagation?
What are the causes of such changes?
Do the in situ properties of ICMEs differ depending on whether they exhibit complexity changes?  
We consider multi-spacecraft observations of 31 ICMEs by MESSENGER, Venus Express, ACE, and STEREO between 2008 and 2014 while radially aligned. By analyzing their magnetic properties at the inner and outer spacecraft, we identify complexity changes which manifest as fundamental alterations or significant re-orientations of the ICME. Plasma and suprathermal electron data at 1~au, and simulations of the solar wind enable us to reconstruct the propagation scenario for each event, and to identify critical factors controlling their evolution. Results show that $\sim$65\% of ICMEs change their complexity between Mercury and 1~au and that interaction with multiple large-scale solar wind structures is the driver of these changes. Furthermore, 71\% of ICMEs observed at large radial ($>$0.4~au) but small longitudinal ($<15^{\circ}$) separations exhibit complexity changes, indicating that propagation over large distances strongly affects ICMEs. Results also suggest ICMEs may be magnetically coherent over angular scales of at least 15$^\circ$, supporting earlier theoretical and observational estimates. This work presents statistical evidence that magnetic complexity changes are consequences of ICME interactions with large-scale solar wind structures, rather than intrinsic to ICME evolution, and that such changes are only partly identifiable from in situ measurements at 1~au.
\end{abstract}



\section{Introduction} 
\label{sec:introduction}


Coronal mass ejections (CMEs) consist of large-scale plasma and magnetic field structures erupted from the Sun into interplanetary space \citep{Webb2012}.
When probed in situ by spacecraft monitoring the conditions of the interplanetary medium, they are referred to as interplanetary CMEs (hereafter ICMEs) and appear as magnetically-dominated (i.e. low plasma $\beta$, that is the ratio of the plasma pressure to the magnetic pressure) structures often exhibiting smoothly-rotating magnetic fields, which are typically interpreted as flux-rope (FR) structures with a helical magnetic field wrapping around a central axis \citep{Klein1982, Kilpua2017}.
Additionally, ICMEs that are faster than the local magnetosonic speed in the solar wind reference frame drive forward shocks and sheaths \citep{Richardson2010, Kilpua2017, Jian2018}.
ICMEs are observed passing over Earth at an average rate of 1--2 per month \citep{Richardson2010}, where, together with their shocks and sheaths, they are the main drivers of strong geomagnetic storms \citep{Gosling1991, Zhang2007, Lugaz2016, Kilpua2017, Kilpua2019}.

Expanding on earlier datasets, recent cruise phase and orbital data from heliospheric and planetary missions has shed new light on ICME characteristics at different heliocentric distances \citep[e.g.][]{Liu2005, Ebert2009, Winslow2015, Good2016, Lee2017, Davies2021b}, revealing different parameter properties from those observed near Earth's orbit, which can also be extremely variable in space and time.
In fact, the evolution of ICMEs during propagation through interplanetary space is shaped by the interplay of internal and external factors controlling the interaction of ICMEs with the surrounding solar wind and other transients therein \citep{Manchester2017}. 
These manifest into four main forms: 
(1) ICME expansion, which controls its size, internal density, pressure, and magnetic field magnitude \citep{Demoulin2009}; 
(2) the interaction with the surrounding solar wind via drag forces, which controls the ICME kinematic properties \citep[][]{Cargill2004, Vrsnak2010};
(3) other forms of energy/momentum exchanges due to the interaction with, e.g., interplanetary shocks driven by various interplanetary structures, which affects the thermal, magnetic, kinematic, and size properties of ICMEs \citep[e.g.][]{Lugaz2015, Zhuang2019, Scolini2020};
and (4) magnetic reconnection phenomena occurring at ICME boundaries \citep[i.e. magnetic erosion and flux injection;][]{Dasso2006, Ruffenach2012} or in their interiors \citep{Crooker1998}, which can alter the connectivity, topology, and size of ICME magnetic structures.

Each of these phenomena contributes to the ultimate evolutionary path of individual ICMEs, but the degree of their influence on the evolution of large-scale ICME properties greatly depends on the ambient solar wind conditions through which an ICME propagates.
Observational and modeling studies established that during propagation, ICMEs undergo a number of large-scale structural changes which include kinks, front flattening \citep[e.g.][]{Savani2011, Davies2021}, rotations and deflections of ICME magnetic structures \citep[e.g.][]{Isavnin2014, Wang2014, Kay2015}, deformations of their front convexity \citep[e.g.][]{Odstrcil1999, Manchester2004}, as well as local magnetic field distortions \citep[e.g.][]{Torok2018}. 
Most importantly, all of the aforementioned effects appear amplified by interactions with high-speed streams (HSSs), corotating/stream interaction regions (CIRs/SIRs), the heliospheric current/plasma sheet (HCS/HPS) \citep[e.g.][]{Odstrcil1999b, Odstrcil1999c, Rodriguez2016, Winslow2016, Winslow2021, Zhou2017, Liu2019, Davies2020, Scolini2021}, as well as other ICMEs \citep[e.g.][]{Lugaz2017, Scolini2020}. 

Direct quantification of the radial evolution of ICMEs can only be achieved through high-quality in situ measurements of approximately the same portion of a given ICME at different heliocentric distances via multi-spacecraft crossings, i.e. by radially-aligned spacecraft configurations. However, such observations are difficult to achieve due to the limited number of assets available (in terms of missions and instruments), and the paucity of continuous observations of the solar wind properties, particularly for studies relying on missions orbiting in and out of planetary magnetospheres/atmospheres. 
Overcoming the scarcity of data, statistical studies combining multiple ICME datasets obtained at different heliocentric distances
allowed the characterization of general trends affecting ICME evolution, particularly in the case of events observed by multiple spacecraft in radial alignment \citep{Good2019, Lugaz2020, Salman2020}. 
Nonetheless, such studies have so far only provided an average picture of the evolution of ICMEs during propagation, rather than diving deep into the analysis of individual events necessary to determine the causes behind the non-ideal evolutionary behavior observed in a number of ICMEs (as recently reported by, e.g., \citet{Lugaz2020} and \citet{Winslow2021}). 
Moreover, past statistical studies investigating the evolution of ICME magnetic topologies during propagation have almost entirely focused on the investigation of ICMEs exhibiting classical FR signatures \citep[e.g.][]{Good2019}, whose properties can be investigated through numerous in situ fitting techniques \cite[e.g.][]{AlHaddad2013}, and neglected non-flux rope configurations, which although more difficult to characterize, are nevertheless frequently observed at 1~au \citep{NievesChinchilla2019} \citep[for a notable exception, see][]{Lugaz2020b}.
As discussed in the following sections, complex magnetic configurations within ICMEs are in fact often the result of magnetic complexity changes attributable to the interaction of ICMEs with other solar wind structures, while classical FR structures are often a proxy for unperturbed propagation. Given their bias towards ideal ICME magnetic structures, it is therefore hardly surprising that previous studies often highlighted little changes in the FR properties between different observing spacecraft, suggesting a propagation scenario compatible with self-similar evolution.

In-depth studies of individual ICMEs observed by multiple spacecraft during periods of radial alignment have been extremely insightful for our understanding of the various phenomena controlling the evolution of their magnetic structures. Such studies have showcased a wide variety of evolutionary behavior, ranging from essentially self-similar \citep{Nakwacki2011, Moestl2012, Good2015, Good2018} to strongly atypical \citep{Leitner2007, NievesChinchilla2012, Winslow2016, Winslow2021, Vrsnak2019, Weiss2021}, posing questions on the frequency and causes of such a large variation of evolutionary trends. 
Advances towards a more comprehensive and fundamental understanding of the radial variations affecting ICMEs have come from tailored investigations of magnetic complexity changes for a limited set of case studies \citep{Winslow2016, Winslow2021, Winslow2021b}, all of which stressed the role of solar wind interplanetary structures as drivers of magnetic complexity changes observed within ICMEs. Yet, the limited number of events considered provided only a partial picture of such phenomenon, necessitating a more extensive investigation.
We also point out that a careful consideration of the effect of different spacecraft trajectories through a given ICME structure, particularly in the case of non-perfect radial alignments, is critical to the interpretation of multi-spacecraft ICME observations at different heliocentric distances \citep[e.g.][]{Lugaz2018}. In this respect, previous studies also suggested that highly inclined magnetic flux ropes might appear very different even when observed at $\sim5^\circ$ in longitudinal separation \citep[e.g.][]{Kilpua2009}, while in comparison, low-inclination flux ropes might appear highly coherent over similar angular scales \citep[e.g.][]{Davies2021}. To what extent the inclination of ICME magnetic flux ropes influences magnetic complexity changes and estimates of the scale of magnetic coherence \citep{Owens2017, Owens2020, Lugaz2018} within ICMEs observed by radially-aligned spacecraft is currently an open question.

This work is motivated by the need to further explore the relationship between large-scale changes affecting ICMEs, and interactions with solar wind structures, particularly HSSs, CIRs/SIRs, and the HCS/HPS, beyond individual case studies. In particular, we perform a statistical analysis of magnetic complexity changes occurring within ICMEs observed by multiple inner heliospheric spacecraft in radial alignment, with the aim of answering the following science questions: 
(1) How frequently do ICMEs undergo magnetic complexity changes during propagation through interplanetary space? 
(2) What are the causes of such changes?
(3) Do the in situ properties of ICMEs differ depending on whether they exhibit complexity changes during propagation?  

We tackle the questions above by analyzing a statistical set of 31 ICMEs that were observed in situ by multiple spacecraft at different heliocentric distances between Mercury's orbit and 1~au in the period 2008--2014. Compared to previous studies, we do not restrict our investigation to ICMEs exhibiting FR magnetic signatures, which allows us to consider a higher number of events and achieve statistical results. We also use in situ observations to identify any intervening structure in the ambient solar wind that might have affected the propagation of the ICMEs of interest. In addition, to place any observed changes in ICME magnetic properties into context, we use simulations of the ambient solar wind obtained via the WSA-ENLIL model hosted at the NASA Community Coordinated Modeling Center (CCMC; \url{https://ccmc.gsfc.nasa.gov}) to identify additional interactions with solar wind structures that might have gone undetected in the available in situ data.
%

The paper is structured as follows.
In Section~\ref{sec:data}, we describe the event selection procedure and the in situ plasma and magnetic field data used to identify the boundaries of ICME structures observed in situ at different spacecraft.
In Section~\ref{sec:methods}, we introduce the methodology used to classify ICME magnetic signatures, we define the measures of complexity changes between observing spacecraft, and we elaborate on the identification of interplanetary structures that interacted with the ICMEs during their propagation from the inner to the outer spacecraft.
In Section~\ref{sec:results}, we present the results of our statistical analysis and interpret them in light of available observations, theoretical arguments, and numerical evidences.
Finally, in Section~\ref{sec:conclusions}, we summarize the main findings and present our conclusions. 

\section{Event Selection and Boundary Identification}
\label{sec:data}

We select events based on the ICME catalog recently compiled by \citet{Salman2020}, which includes 47 ICMEs measured in situ by two or three radially-aligned spacecraft (i.e. MESSENGER, Venus Express, STEREO, Wind/ACE) between 2008 and 2014, and in addition to the ICME in situ properties, it also contains information on their CME counterparts and kinematic properties close to the Sun as measured by coronagraphs. ICMEs were selected according to the following criteria: 
each event had to be observed by at least two spacecraft in radial alignment.
The longitudinal separation between observing spacecraft had to be less than $35^\circ$, so to increase the likelihood of two or more radially aligned spacecraft observing both the sheath and the ejecta, rather than only the sheath itself \citep[][]{Good2016}. Yet, for more than two-thirds of the ICMEs, longitudinal separations between the measuring spacecraft were within $20^\circ$.  
Coronagraphic observations were used to identify CME counterparts, as inferred from the agreement between the CME launch direction and the positioning of the corresponding spacecraft during the event time interval. Additionally, in situ timings were verified against arrival times predicted from coronagraphic observations with the help of the drag-based propagation model \citep[DBM;][]{Vrsnak2013}, to identify actual multi-spacecraft events. 
As also mentioned by \citet{Salman2020}, some of the CME--ICME associations were difficult to verify for all instances, with specific cases having multiple possible CME counterparts or no clear CME counterparts. To further support CME--ICME associations based on coronagraphic and in situ data, in this paper, we have searched for their heliospheric counterparts as observed by the STEREO Heliospheric Imager \citep[HI;][]{Harrison2005} and listed in available databases, in particular the Heliospheric Cataloguing, Analysis and Techniques Service \citep[][\url{https://doi.org/10.6084/m9.figshare.5803152.v1}]{HELCATSWP2} catalog \citep{Harrison2018}. 
Heliospheric CME counterparts from the HELCATS catalog are provided in Table~S1 (included as supporting information) and in Table~\ref{tab:event_list_sample} (providing a sample) for each ICME considered in this study. 
For some of the events, the association between the coronal and HI CME counterparts, and the ICMEs detected in situ could be confirmed against the HELCATS WP4 LINKCAT catalog \citep[][\url{https://doi.org/10.6084/m9.figshare.4588330.v2}]{HELCATSWP4LINKCAT}. 
However, due to the difficulty in analyzing HI data and the fact that HI detectors are designed to function better for Earth-directed CMEs, we found that not all CME--ICME pairs were tracked by heliospheric imagers, and that the information provided by such data products should not be considered as definitive in improving the accuracy of the CME--ICME associations in all cases.

We analyze each event from this catalog using in situ data available at the locations of observation.
To investigate ICMEs at 1~au, we use 1-sec magnetic field data from the magnetometer \citep[MAG;][]{Smith1998} and 64-sec plasma data from the Solar Wind Electron, Proton and Alpha Monitor \citep[SWEPAM;][]{McComas1998}, on board the Advanced Composition Explorer (ACE) mission \citep{Stone1998} orbiting the Sun–Earth Lagrange 1 (L1) point; we also use 1-sec magnetic field data from the magnetometer \citep[MAG;][]{Acuna2008}, and 1-min plasma data from the In situ Measurements of Particles And CME Transients \citep[IMPACT;][]{Luhmann2008} and the Plasma And Suprathermal Ion Composition \citep[PLASTIC;][]{Galvin2008} on board the Solar TErrestrial RElations Observatory-Ahead (STEREO-A) and -Behind (STEREO-B) twin spacecraft \citep{Kaiser2008}. 
Additionally, we use electron pitch angle distribution (PAD) data from ACE/SWEPAM at 272~eV (\url{http://www.srl.caltech.edu/ACE/ASC/DATA/level3/swepam/data/}) and from the STEREO/IMPACT Solar Wind Electron Analyzer \citep[SWEA;][\url{http://stereo.irap.omp.eu/CEF/PAD/}]{Sauvaud2008} at energies above $\sim 100$~eV, i.e. in the suprathermal energy range \citep[][]{Shodhan2000, Zurbuchen2006}.
At Mercury, orbiting at heliocentric distances between 0.31~au and 0.46~au, we make use of 0.5-sec high-resolution magnetic field data provided by the magnetometer \citep[MAG;][]{Anderson2007} on board the MErcury Surface, Space ENvironment, GEochemistry, and Ranging \citep[MESSENGER;][]{Solomon2007} mission. 
Finally, at Venus, located at 0.7~au from the Sun, we make use of 4-sec magnetic field data provided by the magnetometer \citep[MAG;][]{Zhang2006} on board the Venus Express \citep[VEx;][]{Titov2006} mission. 
Magnetic field data obtained by different spacecraft are converted to radial-tangential-normal (RTN) coordinates in order to compare signatures at different spacecraft.

In this work, we use the term ``magnetic ejecta'' (ME) in a similar manner to the way it was used by \citet{Winslow2015} and \citet{Salman2020} to refer to the magnetically-dominated portion of the ICME. This is because for observations at MESSENGER and VEx, no plasma observations are available, and therefore it cannot be verified whether ICMEs also exhibited the expected density, plasma $\beta$, and temperature decreases associated with magnetic clouds \citep[MCs;][]{Burlaga1981}. 
In this study, we therefore identify as ME that part of an ICME exhibiting enhanced magnetic field and low levels of magnetic fluctuations compared to the preceding and following interplanetary magnetic field.
We also note that to determine the ME boundaries, we started from the boundaries listed in existing catalogs compiled by
\citet{Winslow2015}, \citet{Good2016}, \citet{NievesChinchilla2018} (\url{https://wind.nasa.gov/ICME_catalog/ICME_catalog_viewer.php}), and
\citet{Jian2018} (\url{https://stereo-ssc.nascom.nasa.gov/data/ins_data/impact/level3/ICMEs.pdf}). We further modified some of these boundaries to better reflect our ME identification criteria, i.e. of enhanced magnetic field and low levels of fluctuations, as documented in Table~S1 provided as supporting information. 
A sample is provided in Table~\ref{tab:event_list_sample}.

\begin{splitdeluxetable*}{lccccccccBccccccccBccccccc}
\tablecaption{Sample of the ICMEs considered in this work. 
Each event is labeled by a unique number.
The inner and outer observing spacecraft (referred to as ``SC1'' and ``SC2'' in brief) and their longitudinal separation ($\Delta \phi$) are listed.
``M'' stands for MESSENGER, ``STA'' for STEREO-A, and ``STB'' for STEREO-B.
The timings at both spacecraft include: the shock/discontinuity time (if present), the ME start and end times, and the FR start and end times (if present). 
The heliocentric distance of the inner and outer spacecraft ($r_1$ and $r_2$) are also provided.
The FR/ME class observed at each observing spacecraft, together with the index $\Delta \mathbb{C}$ indicating the presence or absence of complexity changes detected between the inner and outer spacecraft, are indicated.
The final six columns list the interactions with various solar wind structures, the presence or absence of BDEs during the FR (when present) or ME passage, 
and the ID(s) of the heliospheric CME counterparts from the HELCATS catalog.}
\label{tab:event_list_sample}
\tablewidth{0pt}
\tablehead{
\colhead{\#} & \colhead{SC1--SC2} & \colhead{$\Delta \phi$} & \multicolumn{5}{c}{SC1 Timing [yyyy-mm-dd HH:MM]} & 
\colhead{$r_1$} & \multicolumn{5}{c}{SC2 Timing [yyyy-mm-dd HH:MM]}& \colhead{$r_2$} 
& \colhead{FR/ME class} & \colhead{FR/ME class} & \colhead{$\Delta \mathbb{C}$} 
& \colhead{Interaction} & \colhead{Interaction} & \colhead{Interaction} & \colhead{Interaction} & \colhead{BDEs} & \colhead{HELCATS} \\
\colhead{} & \colhead{Name} & \colhead{$[^{\circ}]$} & \colhead{Shock/Discontinuity} & \colhead{ME start} & \colhead{ME end} & \colhead{FR start} & \colhead{FR end} & \colhead{[au]} & \colhead{Shock/Discontinuity} & \colhead{ME start} & \colhead{ME end} & \colhead{FR start} & \colhead{FR end} & \colhead{[au]} & \colhead{at SC1} & \colhead{at SC2} & \colhead{} & \colhead{with HSS} & \colhead{with SIR} & \colhead{with HCS} & \colhead{with shock} & \colhead{within FR/ME}
 & \colhead{event ID(s)} 
}
\decimals
\startdata
1 & M--VEx & $1.6$ & 
2011-10-15 08:27 & 2011-10-15 11:14 & 2011-10-16 06:43 & 2011-10-15 11:14 & 2011-10-16 06:43 & 0.46 &
2011-10-16 00:50 & 2011-10-16 06:06 & 2011-10-17 09:39 & 2011-10-16 06:06 & 2011-10-17 09:39 & 0.73 & 	
$Fr$ & $F^-$ & 0 & Probable & No & No & No & Not available & {Not observed} \\
2 & M--VEx & $1.1$ & 
2012-03-07 04:38 & 2012-03-07 06:11 & 2012-03-07 18:01 & -- & -- & 0.32 &
2012-03-07 13:26 & 2012-03-07 20:14 & 2012-03-08 11:43 & -- & -- & 0.72 & 	
$E$ & $E$ & 0 & No & Probable & No & No & Not available & {\texttt{HCME\_A\_\_20120307\_01}, \texttt{HCME\_B\_\_20120307\_01}} \\
4 & M--STB & $4.8$ & 
2011-11-04 15:09 & 2011-11-05 00:45 & 2011-11-05 15:41 & 2011-11-05 00:45 & 2011-11-05 15:41 & 0.44 &
2011-11-06 05:11 & 2011-11-06 22:50 & 2011-11-09 04:00 & 2011-11-06 22:50 & 2011-11-08 13:00 & 1.09 & 	
$Fr$ & $Fr$ & 1 & Probable & No & No & No & Partial & {Not observed} \\
5 & M--STA & $4.6$ & 
2011-12-30 16:27 & 2011-12-30 21:00 & 2011-12-31 09:20 & 2011-12-30 21:00 & 2011-12-31 09:20 & 0.42 &
2012-01-01 13:22 & 2012-01-01 17:00 & 2012-01-02 04:00 & 2012-01-01 17:00 & 2012-01-02 04:00 & 0.96	&
$Fr$ & $F^+$ & 1 & Yes & Yes & Yes & No & No & {\texttt{HCME\_A\_\_20111229\_02}} \\
16 & VEx--STB & $8.5$ & 
2008-12-29 20:46 & 2008-12-29 20:46 & 2008-12-30 04:58 & 2008-12-29 20:46 & 2008-12-30 04:58 & 0.72 &
2008-12-31 02:00 & 2008-12-31 02:00 & 2009-01-01 07:20 & 2008-12-31 04:00 & 2009-01-01 01:00 & 1.03 &
$Fr$ & $Fr$ & 0 & No & No & No & No & No & {\texttt{HCME\_A\_\_20081227\_01}} \\
17 & VEx--STA & $9.2$ & 
2009-06-02 18:39 & 2009-06-02 18:39 & 2009-06-03 12:20 & 2009-06-02 18:39 & 2009-06-03 12:20 & 0.73 &
2009-06-03 00:00 & 2009-06-03 00:00 & 2009-06-04 22:40 & -- & -- & 0.96 & 
$Fr$ & $C_x$ & 1 & No & No & No & Yes & Partial & {\texttt{HCME\_B\_\_20090530\_01}} \\
\enddata
\tablecomments{Table \ref{tab:event_list_sample} is provided in its entirety as supplementary material. 
A sample is presented here for guidance regarding its form and content.}
\end{splitdeluxetable*}

\begin{figure*}
\centering
{\includegraphics[width=1\hsize]{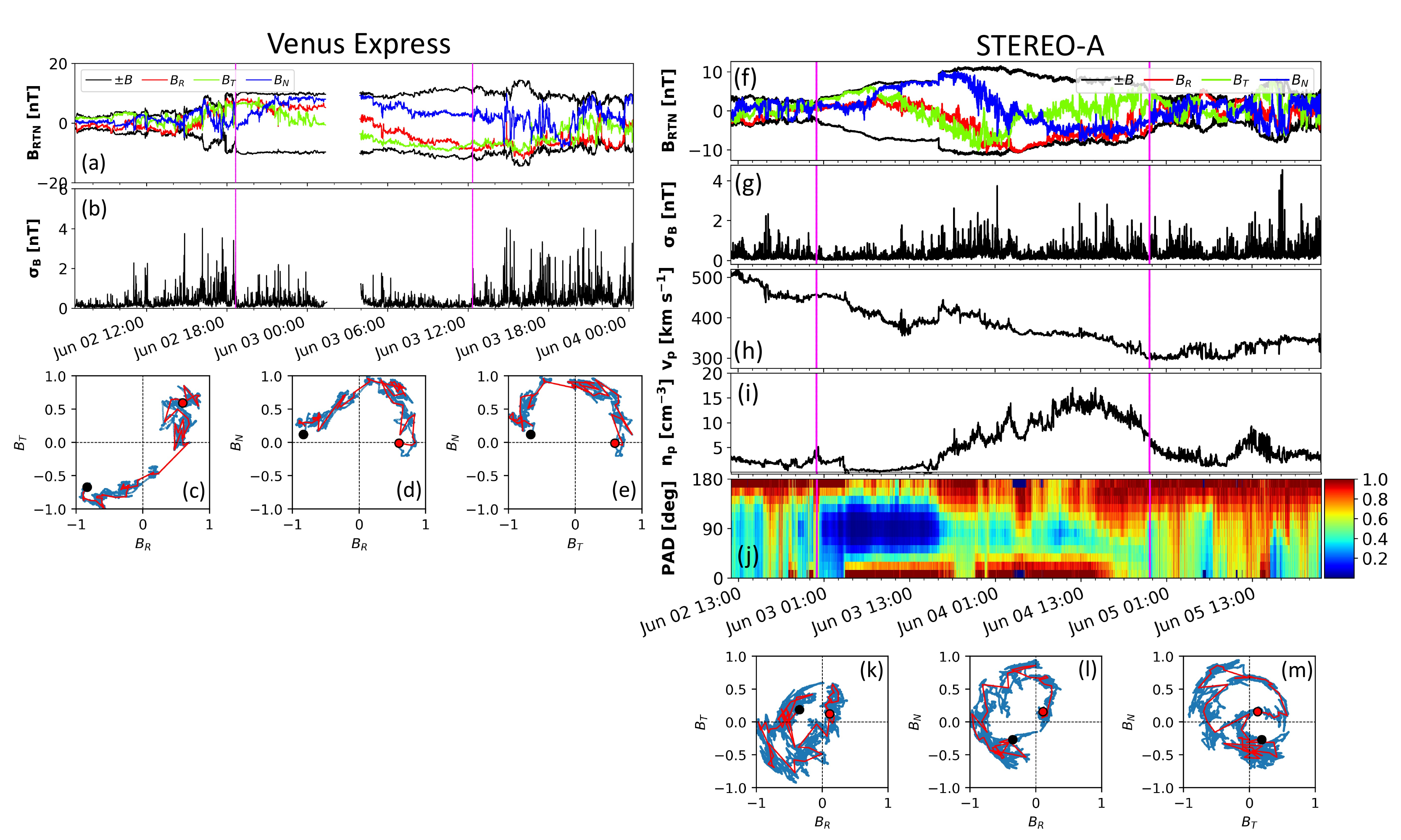}}
\caption{Identification of ICMEs from magnetic field time series for Event~17 in Table~S1, arriving at VEx and STEREO-A on June 2 and 3, 2009, respectively. In this case, the two observing spacecraft were separated by $9^\circ$ in longitude and 0.23~au in the radial direction.
(a)--(e): signatures at VEx. (f)--(m): signatures at STEREO-A.
Panels (a), (f) show the magnetic field time series, and panels (b), (g) show the magnetic field standard deviation. 
Panels (h), (i), and (j) shows the speed, and the proton number density, and the normalized suprathermal electron pitch angle distribution data detected at STEREO~A. 
The vertical solid lines mark the FR boundaries.
Panels (c)--(e) and (k)--(m) show the magnetic hodograms for the FR/ME periods.
An $Fr$ class is identified at VEx, while a $C_x$ class is visible at STEREO-A.} 
\label{fig:figure1}
\end{figure*}

In addition to the ME boundaries, for each event we determine whether the ME or a significant portion of it (lasting more than $\sim25-30$\% of the ME duration) exhibited smooth rotations of the magnetic field component, i.e. signatures compatible with an FR configuration. Hereafter, we refer to these as FRs. We find that some of the events have no FR within the ME, some have an FR that extends throughout the whole ME, and some have an FR that only covers a portion of the ME. If present, the FR portion of the ME is used as a reference to identify complexity changes affecting MEs during propagation because it allows us to identify finer structural changes including rotations, e.g., by means of in situ fitting techniques (see Section~\ref{subsubsec:complexity_change_def_B}). 
If no FR is present within the ME, we consider the whole ME in order to evaluate complexity changes.
An example of this boundary identification procedure is provided in Figure~\ref{fig:figure1} for event~17 observed at VEx and STEREO-A. A discussion on the relationship between ME and FR boundaries is provided in Section~\ref{subsubsec:duration_scaling}.

For the purposes of this paper, we further remove from the initial list all the events that have significant data gaps, i.e. gaps extending more than half of the ME duration, and for which the available data show no obvious rotation in the magnetic field components. We also discard all the events associated with interacting ICMEs, i.e. those events for which an interaction among multiple ICMEs is recognizable in situ, as the investigation of ICME--ICME interactions goes beyond the scope of this study. This results in a total of 31 suitable ICMEs identified, while 16 events were discarded from our analysis.
Each event was observed at two spacecraft (i.e. the inner spacecraft is either MESSENGER or VEx, and the outer spacecraft is either VEx, ACE, STEREO-A or STEREO-B) during periods of radial alignment, with longitudinal separations among the observing spacecraft ranging between 1$^\circ$ and 32$^\circ$.
The 31 ICMEs used in this study, including the details of spacecraft alignments and arrival times at each spacecraft, are listed in Table~S1. 
Further details about this database are given later on in the text (see Section~\ref{sec:results}).

\section{Methods}
\label{sec:methods}

\subsection{Definition of Magnetic Complexity Changes}
\label{subsec:complexity_changes}

In the following subsections, we formalize the definition of ICME magnetic complexity changes which will be used throughout this work to identify major alterations in the magnetic configuration of ICMEs detected between two spacecraft, in a manner that is as unbiased as possible. As detailed in Sections~\ref{subsubsec:complexity_change_def_A} and \ref{subsubsec:complexity_change_def_B} below, we consider two major indicators of magnetic complexity changes within ICMEs: 
(1) fundamental alterations of the ICME magnetic structure as indicated by the change in FR (when present) or ME category detected between the two observing spacecraft; 
and (2) significant re-orientations of ICME structures as inferred from magnetic hodograms and in situ fitting techniques. 
We summarize the evolution of magnetic complexity of each ICME using the index $\Delta \mathbb{C}$, 
which assumes a value of $\Delta \mathbb{C}=0$ in the case of events that do not change their magnetic complexity between the inner and outer spacecraft, and of $\Delta \mathbb{C}=1$ in the case of events that do change their magnetic complexity between the inner and outer spacecraft.

\subsubsection{Condition A: Fundamental Alteration of FR/ME Structures}
\label{subsubsec:complexity_change_def_A}

We construct the magnetic hodograms for each ICME FR (when present) or ME in our list, and visually classify their internal magnetic structure at a given spacecraft using a morphological classification scheme based on the one proposed by \citet{NievesChinchilla2018, NievesChinchilla2019}. 
This classification sorts FRs/MEs into five classes, which encompass the wide variety of signatures observed: 
$Fr$, exhibiting a maximum rotation in any magnetic component close to $180^\circ$;
$F^-$, with maximum rotation in any magnetic component smaller than $120^\circ$;
$F^+$, with maximum rotation in any magnetic component larger than $240^\circ$;
$C_x$, complex structures with more than one rotation, e.g. with different radii of curvature and/or different rotation directions; 
and $E$, characterized by a lack of clear rotations. An example of $Fr$ and $C_x$ signatures are shown in Figure~\ref{fig:figure1}.
Such a classification allows the categorization of FR/ME signatures without imposing a priori restrictions on them other than having a high magnetic field with low levels of fluctuations. Nevertheless, this classification developed from the assumption that the internal magnetic structure of ICMEs in their lowest energy state can be locally described as a single FR with a helical magnetic field wrapped around a central axis. We remark that this idea has been debated and contrasted in previous works \citep{Owens2016, AlHaddad2019}. It also remains unclear whether any individual FR model, among the many that have been developed  \citep[e.g.][]{Burlaga1988, Lepping1990, Hidalgo2002a, Hidalgo2002b, Isavnin2016, NievesChinchilla2016, NievesChinchilla2018b, Moestl2018}, is sufficient to describe the observed variety of FR and ME signatures, including $F^+$, $E$, and $C_x$ classes.

In the aforementioned classification scheme, all $F$ types are compatible with crossings through single FR structures.
In the following analysis, we consider $Fr$ and $F^-$ classes as a single category, since rotations close to and smaller than $180^\circ$ in the magnetic field components are both compatible with different spacecraft trajectories through a helical magnetic field wrapped around an axis, e.g. arising from small and large impact parameter crossings, respectively \citep{NievesChinchilla2019}. 
Magnetic field rotations significantly larger than $180^\circ$ ($F^+$ class) can be interpreted as signatures of FRs with significant curvature, e.g. spheromaks \citep{Scolini2021}, or potentially double FRs \citep{Lugaz2013}, and are therefore kept as a separate category for the sake of clarity. Neither of the remaining two classes ($C_x$ and $E$) can be reproduced by an FR model, and due to their distinct characteristics, they are maintained as separate categories and considered as indicative of crossings through fundamentally different magnetic structures. 
We note that \citet{Owens2016} argued that some ICME legs might actually have non-flux rope configurations such as untwisted magnetic field lines. In this scenario, $E$ types may also be consistent with crossings through untwisted legs carrying little-to-no magnetic field rotation.

For each ICME, we then compare the classifications obtained at the inner and outer spacecraft, and use them as prime indicators to assess whether a given ICME has undergone a magnetic complexity change during propagation, i.e. we define a complexity change as any change in the FR/ME category detected between the two observing spacecraft.

\subsubsection{Condition B: Re-orientation of FR/ME Structures}
\label{subsubsec:complexity_change_def_B}

We further scrutinize events that retained their magnetic configuration as defined in Section~\ref{subsubsec:complexity_change_def_A} above, searching for complexity changes that manifest as significant re-orientations (i.e. longitudinal and latitudinal rotations) of FR/ME structures.

For those events that are observed as $Fr$/$F^-$ types at both spacecraft, we fit observational data using a linear force-free (LFF) fitting model based on the force-free constant-$\alpha$ FR model developed by \citet{Burlaga1988} and subsequently optimized by \citet{Lepping1990}.
\citet{Lepping2003} showed that different noise levels can affect the results of the LFF fitting technique leading to uncertainties up to $\pm 13^\circ$ and $\pm 30^\circ$ in the reconstructed latitudinal and longitudinal FR axial directions, while \citet{AlHaddad2013, AlHaddad2018} assessed that different fitting models can differ up to $\pm 45^\circ$ in latitude and/or $\pm 60^\circ$ in longitude in the reconstructed orientation of the FR axis.
Since we use a single fitting technique, we take a conservative approach and consider changes in the reconstructed FR axis direction that are greater than $\pm 45^\circ$ in latitude and/or $\pm 60^\circ$ in longitude as indicators of complexity changes having occurred during propagation between the two spacecraft. These thresholds have been chosen based on the results obtained by \citet{AlHaddad2013, AlHaddad2018}, who showed that larger differences in latitude/longitude are most often indicative of actual structure re-orientations rather than due to the particular model used.

In a similar way, events that are observed as $F^+$, $E$, or $C_x$ types at both spacecraft are further scrutinized, and complexity changes are identified when the magnetic hodograms exhibit significant changes in the magnetic field polarities between the two spacecraft, i.e. corresponding to drastic rotations of the ME.

\subsection{Identifying the Drivers of Magnetic Complexity Changes}
\label{subsec:causes_complexity_changes}
To identify the causes of magnetic complexity changes, we search for possible interactions of each ICME with other solar wind structures, namely HSSs, SIRs, the HCS/HPS, as well as isolated interplanetary shocks, that might have occurred during propagation between the two observing spacecraft. To do so, we first investigate the presence of solar wind structures interacting with each ICME based on in situ plasma and magnetic field data available at each spacecraft. We then cross-check our identifications with publicly available catalogs, particularly the Heliospheric Shock Database, generated and maintained at the University of Helsinki \citep[\url{http://ipshocks.fi}]{Kilpua2015}, the Level~3 STEREO IMPACT/PLASTIC ICME, SIR and interplanetary shock lists \citep[\url{https://stereo-dev.epss.ucla.edu/l3_events}]{Jian2013, Jian2018, Jian2019}, and the Near-Earth ICME catalog by \citet{Cane2003} and \citet{Richardson2010} (\url{http://www.srl.caltech.edu/ACE/ASC/DATA/level3/icmetable2.htm}). We note that in in situ data, we identify HSSs as regions of $v_r \ge 500$~km/s, or more generally as solar wind structures propagating faster than the a given ICME at 1~au \citep{Cranmer2017}.
In the rest of this paper, we refer to in situ evidence of interactions between ICMEs and solar wind structures as ``confirmed'' interactions.

To account for additional interactions that might have started and ended between the ICME passages at the inner spacecraft and outer spacecraft, we check WSA-ENLIL simulations of the ambient solar wind available on the NASA CCMC server (\url{https://ccmc.gsfc.nasa.gov}; keywords: \texttt{Dan\_Aksim, YYYY~amb~(from restart)}, where \texttt{YYYY} is the year of interest) to verify the presence of other solar wind structures that might have interacted with the ICME between the two spacecraft encounters.
In particular, we look at the presence of solar wind interplanetary structures between the two observing spacecraft in the period between the ICME arrival at the inner observing spacecraft, and the ICME end at the outer observing spacecraft. We remark that typical ICME propagation times between the inner and outer spacecraft are around 1 day for MESSENGER--VEx and VEx--1~au conjunctions, and around 2 days for MESSENGER--1~au conjunctions.
In WSA-ENLIL simulations, we identify HSSs as regions of $v_r \ge 500$~km/s, and SIRs as dense regions generated by the interaction of fast and slow solar wind, characterized by a scaled number density $n (r \,\, [\mathrm{au}]/1\,\, \mathrm{au})^2$ larger than $\sim 10$~cm$^{-3}$ \citep{Jian2019}.
We note that in WSA-ENLIL simulations, the modeled arrival times of HSSs at given spacecraft locations are not always consistent with the arrival times observed by spacecraft in situ \citep[e.g.][]{Gressl2014}. In this study, we use WSA-ENLIL results as a guide on the global structure of the ambient wind, working under the assumption that HSS arrival times modeled by WSA-ENLIL are approximately correct. 
Therefore, in what follows, we label interactions between ICMEs and solar wind structures inferred from WSA-ENLIL results as ``probable'' interactions.

\section{Results and Discussion}
\label{sec:results}

\subsection{Frequency of FR/ME Types and Complexity Changes}
\label{subsec:results_ME_types}

\begin{figure*}
\centering
{\includegraphics[width=1\hsize]{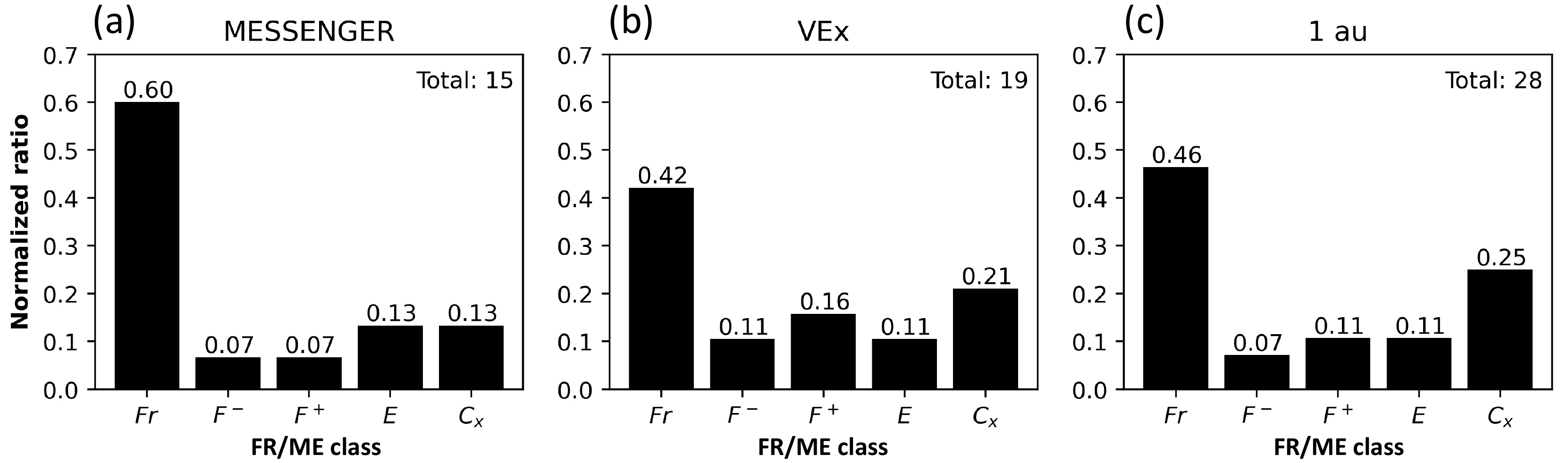}}
\caption{Distribution of FR/ME types observed by spacecraft at MESSENGER (a), VEx (b), and 1~au (c).} 
\label{fig:figure2}
\end{figure*}

As listed in Table~S1, our database consists of a total of 31 ICMEs, each observed at two spacecraft, for a total of 62 observations. 
Table~S1 also reports, for each event, the morphological classification assigned to the FR (when present) or ME magnetic structure at each observing spacecraft. The index $\Delta \mathbb{C}$ is used to indicate an unchanged ($\Delta \mathbb{C} = 0$) or changed ($\Delta \mathbb{C} = 1$) complexity as detected between the two observing spacecraft. We note that in the latter column, the index $\Delta \mathbb{C}$ has been determined by applying both Conditions~A (Section~\ref{subsubsec:complexity_change_def_A}) and B (Section~\ref{subsubsec:complexity_change_def_B}) to each event. The results of the LFF fits for those events that maintained an $Fr/F^-$ category at both spacecraft are provided in Table~\ref{tab:lfffits} for completeness.

Overall, the categorization of MEs/FRs yields the following: 35 $Fr$/$F^-$ types ($\sim56\%$), 7 $F^+$ types ($\sim11\%$), 7 $E$ types ($\sim11\%$), and 13 $C_x$ types ($\sim21\%$) (shown in Figure~\ref{fig:figure2} at MESSENGER, VEx, and 1~au).
However, as visible from Figure~\ref{fig:figure2}, the FR/ME class distribution is significantly more similar between VEx and 1~au than it is at MESSENGER. In particular, at 1~au we observe less $Fr$/$F^-$ (53\%), and more $C_x$ (25\%) types than at MESSENGER (67\% and 13\%, respectively). This simple result is already indicative of a general trend in ICME complexity increase from the inner heliosphere to 1~au. Furthermore, the distribution at 1~au is similar to the one reported by \citet{NievesChinchilla2019} throughout the past two solar cycles, with minor differences (e.g. a higher fraction of $C_x$ events reported in this study) possibly ascribable to solar cycle variations.
For completeness, we report the results from previous studies investigating the fraction of ICMEs having MC signatures at different heliocentric distances.
\citet{Bothmer1996} found that $\sim41\%$ of the fast ICMEs observed by the Helios spacecraft between 1979 and 1981 exhibited MC characteristics. \citet{Richardson2010} reported an approximate fraction of one third of ICMEs showing MC signatures at 1~au. Beyond Earth's orbit, \citet{Rodriguez2004} found $\sim27\%$ ICMEs to be MCs between 1 and 5~au based on Ulysses observations. 
MCs are a subcategory of $Fr/F^-$ types, which explains the lower fractions reported in previous studies compared to those retrieved in this work. Yet, the decreasing trends in the fractions of both $Fr/F^-$ types and MCs detected at increasing heliocentric distances provide independent indications that ICMEs become more complex as they propagate through the inner heliosphere.

Of the 31 events under study, 18 ($\sim 58$\%) did not exhibit fundamental alterations of their internal magnetic configuration as defined using Condition~A, while 13 ($\sim 42$\%, a substantial fraction) did.
Moreover, 7 among the 18 events that did not exhibit fundamental alterations in their magnetic structure, underwent significant re-orientations as determined from  LFF fits and the magnetic hodograms observed at the two observing spacecraft (i.e. as defined by Condition~B, see Table~\ref{tab:lfffits}).
By combining Conditions~A and B together, we conclude that 11 events ($\sim 35$\%) did not change their complexity, while 20 ($\sim 65$\%) did. 
These results show that complexity changes in interplanetary space are much more frequent than estimated by previous studies considering FR ICMEs only \citep[e.g.][]{Good2018, Palmerio2018}, and that they in fact affect the majority of ICMEs observed between Mercury and 1~au.

\begin{table}
\footnotesize
\centering
 \begin{tabular}{c | c c c c c c | c c c c c c | c c} 
 \hline
 \hline
 \multicolumn{1}{c}{}  & \multicolumn{6}{c}{SC1} & \multicolumn{6}{c}{SC2} & \multicolumn{2}{c}{} \\
 Event \# & $H$ & $\theta_0$ ($^\circ$) & $\phi_0$ ($^\circ$) & $y_0$ & $B_0$ (nT)  & $\chi^2_\mathrm{red}$ & $H$ & $\theta_0$ ($^\circ$) & $\phi_0$ ($^\circ$) & $y_0$ &  $B_0$ (nT) & $\chi^2_\mathrm{red}$ & $|\Delta \theta_0|$ ($^\circ$) & $|\Delta \phi_0|$ ($^\circ$) \\
 \hline
 1  & $-1$ & $-2$ & 0 & $-0.70$ & 93 & 0.04     & $-1$ & $-9$ & 1 & $-0.78$ & 50 & 0.02         & 7 & 1 \\
 4  & $-1$ & 16 & 54 & $-0.26$ & 44 & 0.06      & $-1$ & 3 & 166  & $0.65$ & 12 & 0.03          & 13 & \textbf{112}\\
 6  & $1$ & 15 & 92 & $0.26$ & 38 & 0.12       & $1$ & 12 & 14   & $0.53$ & 25 & 0.05          & 2 & \textbf{77} \\
 9 & $-1$ & 6 & 69 & $-0.60$ & 134 & 0.08      & $-1$ & 17 & 152  & $0.12$ & 16 & 0.06         & 11 & \textbf{83} \\
 10 & $1$ & 14 & 326 & $0.64$ & 56 & 0.05       & $1$ & 27 & 279  & $-0.44$ & 7 & 0.03          & 12 & 47 \\
 12 & $1$ & $-3$ & 328  & ${0.79}$ & 64 & 0.03  & $1$ & 17 & 259  & ${-0.71}$ & 21 & 0.05       & 20 & \textbf{69} \\
 14 & $-1$ & $-6$ & 85  & ${-0.09}$ & 47 & 0.04   & $-1$ & $-22$ & 105  & ${-0.04}$ & 14 & 0.04 & 16 & 20 \\
 16 & $-1$ & $-2$ & 167 & $0.87$ & 17 & 0.04    & $-1$ & $-11$ & 138  & $0.59$ & 10 & 0.06      & 9 & 29 \\
 18 & $-1$ & 10 & 36 & $-0.66$ & 19 & 0.06      & $-1$ & $-25$ & 68  & $-0.02$ & 9 & 0.05       & 35 & 32 \\
 20 & $-1$ & 7 & 9 & $-0.92$ & 40 & 0.01         & $-1$ & 4 & 3 & $-0.92$ & 20 & 0.02           & 3 & 6 \\
 22 & $-1$ & 0 & 5 & $-0.78$ & 21 & 0.05         & $-1$ & 1 & 4 & $-0.79$ & 13 & 0.06           & 1 & 1 \\
 23 & $1$ & $-6$ & 312  & $0.23$ & 17 & 0.06    & $1$ & $-25$ & 290 & $0.75$ & 12 & 0.02        & 19 & 22 \\
 25 & $1$ & 57 & 134 & $-0.62$  & 27 & 0.09     & $1$ & $-1$ & 229 & $-0.74$ &  9 & 0.04        & \textbf{58} & \textbf{95} \\
 \hline
 \end{tabular}
 \caption{LFF fit parameters for the events categorized as $Fr$/$F^-$ types at both spacecraft. 
 $H$ -- magnetic chirality, 
 $\theta_0$ -- latitude of FR axis,
 $\phi_0$ -- longitude of FR axis,
 $y_0$ -- impact parameter (normalized by FR radius), 
 $B_0$ -- magnetic field at FR axis, 
 $\chi^2_\mathrm{red}$ -- reduced $\chi^2$.
 $|\Delta \theta_0|$ and $|\Delta \phi_0|$ estimate the FR axis rotation between the two spacecraft.
 The values in bold mark significant re-orientations as defined by Condition~B.}
 \label{tab:lfffits}
\end{table}

\subsubsection{Effect of Spacecraft Radial and Longitudinal Separations}
\label{subsubsec:results_spacecraft_separations}

\begin{figure*}
\centering
{\includegraphics[width=0.55\hsize]{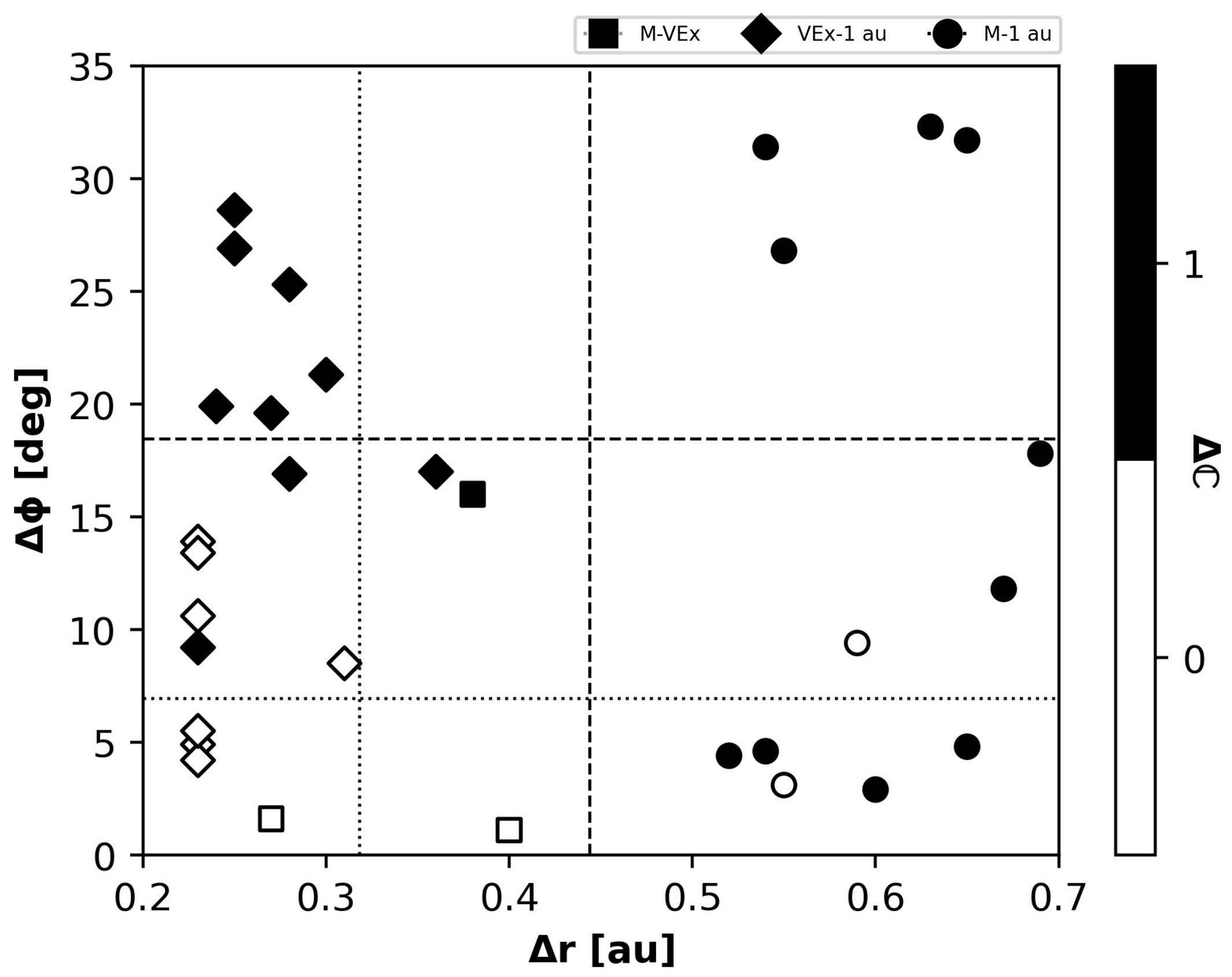}}
\caption{Complexity change $\Delta \mathbb{C}$ as a function of the spacecraft radial ($\Delta r$) and longitudinal ($\Delta \phi$) separations. Different markers are used to indicate different pairs of spacecraft in radial alignment.
The vertical and horizontal lines show the average radial and longitudinal spacecraft separation for events exhibiting (dashed) and not exhibiting (dotted) complexity changes, respectively.}
\label{fig:figure3}
\end{figure*}

Figure~\ref{fig:figure3} shows the complexity evolution of individual ICMEs as a function of the radial and longitudinal separations between the observing spacecraft. ICMEs that preserve their magnetic complexity are observed over shorter spatial and temporal scales than ICMEs that do change their complexity.
We find that unchanged and changed events are observed by spacecraft with mean radial separation of 0.32~au and 0.44~au, respectively.
The mean longitudinal separations for spacecraft observing unchanged and changed events are 7$^\circ$ and 18$^\circ$, respectively.
We further apply the Welch's t-test (i.e. a two-sample location test which is used to test the hypothesis that two populations have equal means) to determine if differences between events that change and that do not change complexity during propagation are statistically significant. We impose a 95\% confidence level, meaning that we can reject the null hypothesis of the test (i.e. that the two populations have equal means) for p-values smaller than 0.05 (5\%).
We find p-values of 0.03 and 0.0001 for radial and longitudinal separations, and conclude that there is a statistically-significant difference in mean values between the two groups. 

In more detail, roughly 82\% (9 out of 11) of the ICMEs that preserve their magnetic complexity are observed by spacecraft in close radial alignment ($\Delta \phi <15^\circ$) and at radial separations $\Delta r \leq 0.4$~au. At such small longitudinal and radial separations, they make up 90\% of the ICMEs (10 out of 11 events).
The remaining $\sim 18\%$ of unchanged events (2 out of 11) are observed in close radial alignment but at larger radial separations than 0.4~au. 
In this region, however, 71\% (5 out of 7) of the ICMEs do change their magnetic complexity, while only 29\% (2 out of 7) do not. As these ICMEs are observed by spacecraft in close radial alignment, and therefore both spacecraft cross through approximately the same region of the ICME, such complexity changes can be most likely attributed to propagation (i.e.\ ``nurture'' effects), rather than to large-scale pre-existing differences in the internal properties within different ME regions.
Oppositely, all of the events observed at smaller radial separations ($\Delta r \leq 0.4$~au) but at larger angular separations ($\Delta \phi > 15^\circ$) exhibit complexity changes, possibly resulting from a combination of initial or pre-existing irregularities in their internal magnetic structure (i.e.\ ``nature'' effects), and propagation effects (due to ``nurture''), which can also include inhomogeneous and variable ambient solar wind conditions through which different parts of an ICME propagate. Complexity changes are also detected in all ICMEs observed at large radial separations ($> 0.4$~au) and large angular separations ($\Delta \phi > 15^\circ$).

We note that for the majority of the events considered, the radial separation between the observing spacecraft also underlies a trend in the spacecraft heliocentric distances, with observations at smaller $\Delta r$ being typically taken at larger distances from the Sun (i.e. at Venus and 1~au) compared to observations at larger $\Delta r$ (i.e. at Mercury and 1~au).
However, as reported by \citet{Scolini2021}, for a given $\Delta r$, a higher probability of developing complexity changes is expected for ICMEs observed by an inner spacecraft located within $\sim0.5$~au. Further observational insights on the radial scale at which ICMEs preserve their magnetic complexity as a function of the heliocentric distance would therefore require additional observations spanning the whole space of possible $r$ and $\Delta r$ parameters.

We also point out that only two out of the 13 events listed in Table~\ref{tab:lfffits} (i.e. events 12 and 25) were observed at angular separations larger than $15^\circ$, and both were associated with complexity changes likely resulting from the interaction with multiple interplanetary structures (as further discussed in Section~\ref{subsec:results_interactions}). This suggests that the significant re-orientations and magnetic complexity changes identified through the application of Condition~B cannot be explained in terms of large-scale geometric effects due to spacecraft crossings through different parts of stretched FRs, as was the case, for example, in the events investigated by \citet{Moestl2012}.
Additionally, the common assumption that an unperturbed ICME magnetic structure should exhibit relatively similar characteristics along different directions implies that drastic magnetic complexity changes as defined by Condition~A cannot be explained only in terms of the angular separations among the spacecraft, particularly for $\Delta \phi$ significantly smaller than the ICME angular width \citep[estimated to be around $\sim40^\circ-60^\circ$ on average;][]{Yashiro2004, Kilpua2011}. 
In such a scenario, the FR/ME category should not change when crossing through the same ICME structure from different directions. 
As this is most probably an over-simplification of the real situation and subject of debate \citep[see e.g.][]{Lugaz2018, Davies2020, Owens2020}, we took a closer look at the events observed above $15^\circ$ of angular separation that exhibited complexity changes as defined by Condition~A. We found that out of 12 events, 4 (33\%; i.e. events 3, 11, 26 and 29) exhibited a complexity change that could not be attributed to any interaction with other solar wind structures (as further discussed in Section~\ref{subsec:results_interactions}). For the remaining 8 cases (66\% of the events), the interaction with interplanetary structures was likely the major contributor to the detected complexity change.
Overall, these arguments suggest that in the large majority of cases, the results reflect actual magnetic complexity changes affecting ICMEs, rather than geometrical effects due to the particular spacecraft trajectories along a given structure.

With respect to the question of whether complexity changes might be significantly affected by the spacecraft trajectory through the ICME magnetic structure depending on its inclination, we also note that in this work, all events that could be fitted using the LFF fitting technique except one (Event~25 at the inner spacecraft) were reconstructed as having a low inclination with respect to the equatorial plane (indicated by the small $\theta_0$ values in Table~\ref{tab:lfffits}). The determination of the inclination of the remaining magnetic structures was complicated by the lack of clear flux rope signatures, and could not be performed. Overall, at least with respect to the subset listed in Table~\ref{tab:lfffits}, the aforementioned results were therefore derived for a population of low-inclination flux ropes, whose signatures appear to be less affected by trajectory effects than highly-inclined flux ropes \citep[e.g.][]{Kilpua2009, Davies2021}. The assessment of possible inclination effects on ICME magnetic complexity changes and magnetic coherence will require the consideration of highly-inclined ICMEs in future studies.

Remarkably, the $\Delta \phi \sim 15^\circ$ breaking point retrieved in this study (i.e., above this longitudinal separation all events exhibit complexity change) is only slightly smaller than previous estimates of the maximum angular width at which ICME magnetic structures behave coherently (i.e. $17^\circ - 26^\circ$ reported by \citet{Owens2020}, and $14^\circ - 20^\circ$ by \citet{Lugaz2018} based on theoretical and observational arguments, respectively). Although making an accurate estimate of the actual scale of ICME coherence is not possible due to the non-negligible radial separations affecting the observations considered in this study, the results in Figure~\ref{fig:figure3} suggest ICMEs might retain some level of coherence over angular scales of the order of $15^\circ$.
We also argue that if multi-point observations with negligible radial separation were available for the events considered, i.e. once any effects related to propagation were removed, coherence would be most probably observed over scales even larger than 15$^\circ$.

Finally, for the 9 events observed in perfect alignment, i.e. with angular separation $<5^\circ$, we report that 6 of them ($\sim67$\%) did not change their FR/ME class, while only 3 ($\sim33$\%) did. These values validate previous conclusions by \citet{Scolini2021} based on the results from numerical simulations (reporting a 69\% to 54\% probability for ICMEs to maintain their FR class, and a 31\% to 46\% probability to change their FR class, depending on the ambient solar wind conditions).
In this case, the average radial separation for ICMEs observed in perfect conjunction is 0.34~au for events that did not change complexity, and 0.58~au for events that did change complexity.

\subsection{Drivers of Magnetic Complexity Changes}
\label{subsec:results_interactions}

\begin{figure*}
\centering
{\includegraphics[width=\hsize]{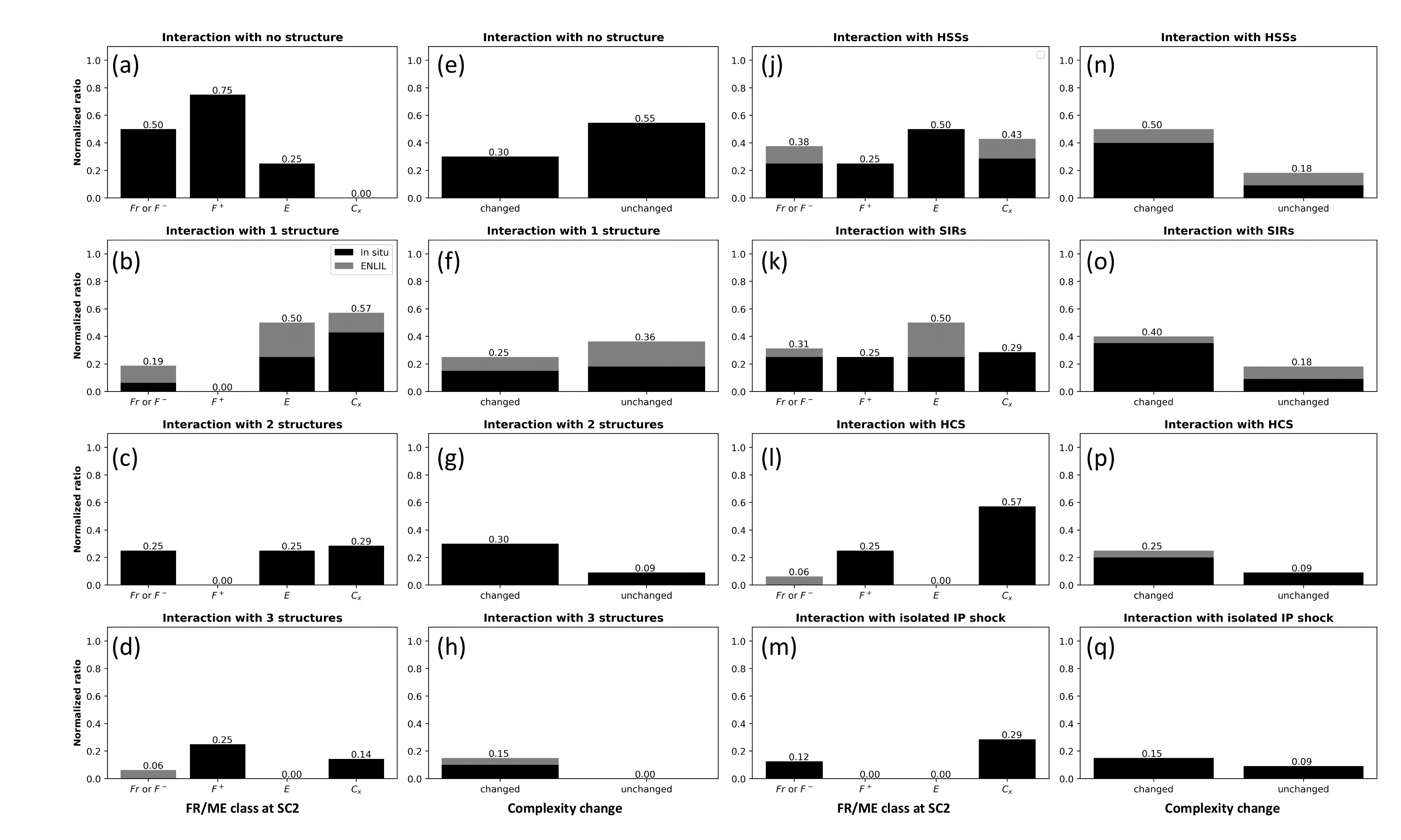}}
\caption{Interaction with solar wind structures as a function of the FR/ME category detected at the outer spacecraft, and of the complexity evolution history.
Black bars indicate confirmed interactions, while grey bars indicate probable interactions based on WSA-ENLIL simulations of the ambient solar wind.}
\label{fig:figure4}
\end{figure*}

As stated in Section~\ref{subsec:results_ME_types}, we found that out of the 31 ICMEs considered, 11 did not change their magnetic complexity, while 20 did. 
Figure~\ref{fig:figure4} reports their probability of interaction with other solar wind structures as a function of the FR/ME category detected at the outer spacecraft, and of the complexity evolution history determined from the analysis described in Section~\ref{subsec:causes_complexity_changes}. In total, we were able to identify {26} confirmed interactions from in situ data, involving {15} ICMEs, and 6 probable interactions from WSA-ENLIL simulations, involving 5 ICMEs. In the following, these two types of interactions are discussed together, while in Figure~\ref{fig:figure4} their contributions are presented in different colors for completeness.
A detailed list of the confirmed and probable interactions identified for each ICME is also included in Table~S1.

We find that {50}\% of $Fr$/$F^-$ types ({8} of 16 events) and 75\% of $F^+$ types (3 of 4 events) do not interact with any structure. This fraction is down to 25\% and 0\% for $E$ types (1 of 4 events) and $C_x$ types (0 of 7 events), respectively (Figure~\ref{fig:figure4}~(a), (b), (c), (d)), indicating that interactions with solar wind structures associate in particular with $E$ and $C_x$ types. As a note of caution, we remark that the $F^+$ and $E$ fractions above are likely to have been influenced by the small number of events observed at outer observing spacecraft (4 for both categories), and might therefore change significantly if more events were available.

Out of the 11 events that do not change their complexity, {6} ({55}\%) do not interact with any structure, while {5} ({45}\%) do interact with one ({4} events, {36}\% of the total) or two (1 event, 9\% of the total) structures (Figure~\ref{fig:figure4}~(e), (f), (g), (h)), suggesting that interaction with solar wind structures does not always induce a change in magnetic complexity. 
We also find that out of the 20 events that do change their complexity, 6 (30\%) do not interact with any structure, while 14 (70\%) do interact with one (25\%, 5 events in total), two ({30}\%, {6} events in total), or three ({15}\%, {3} events in total) structures. We conclude that while an interaction with another structure does not always result in a change in complexity, it does increase the probability for ICMEs to undergo a complexity change. In fact, most of the unchanged events ($\sim 90$\%) either do not interact with any structure, or only interact with a single structure. On the contrary, events that changed their complexity interact with two or more structures much more frequently (in $\sim 45$\% of the cases considered).

When considering the interaction with specific structures (i.e. HSSs, SIRs, HCS, and isolated shocks), we find that $C_x$ types, which have the highest association with interactions, are most often the result of interactions with the HCS (57\%, Figure~\ref{fig:figure4}~(l)). They also have high associations with all other structures (43\% to 29\%, Figure~\ref{fig:figure4}~(j), (k), (m)). The remaining FR/ME categories are more often linked to interactions with HSSs and SIRs (50\% to 25\%), and less often with the HCS (less than 25\%) and isolated shocks (less than 12\% of the cases).

Figure~\ref{fig:figure4}~(n), (o), (p), (q) show that events that change their complexity are more likely to have interacted with any of the structures considered than events that do not change their magnetic configuration. In particular, the primary cause of complexity changes is the interaction with HSSs (found in 50\% of the events exhibiting complexity changes, Figure~\ref{fig:figure4}~(n)), followed by the interaction with SIRs (found in 40\% of the events exhibiting complexity changes, Figure~\ref{fig:figure4}~(o)), HCS (found in {25}\% of the events exhibiting complexity changes, Figure~\ref{fig:figure4}~(p)), and isolated shocks (found in 15\% of the events exhibiting complexity changes, panel n). For comparison, only 18\% of the events do not change their complexity were found to have interacted with HSSs, SIRs (Figure~\ref{fig:figure4}~(n), (o)), and 9\% {with the HCS or} with isolated shocks (Figure~\ref{fig:figure4}~{(p),} (q)).
We note that the correlation between magnetic complexity changes and interactions with solar wind structures presented here has been determined with the aid of in situ magnetic field and (when available) bulk solar wind data, as well as with global simulations of the inner heliospheric solar wind conditions. We investigate in Section~\ref{subsubsec:results_BDEs} below whether additional data products, in particular suprathermal electron PAD data, can provide valuable additional information for the identification of ICME complexity changes.

\subsection{Characteristics of ICMEs Exhibiting Complexity Changes}
\label{subsec:results_insitu_characteristics}

\subsubsection{Bi-directional Electrons}
\label{subsubsec:results_BDEs}

Since the presence of bi-directional suprathermal electron flows (BDEs) is tied to the magnetic topology of ICMEs \citep[e.g.][]{Gosling1987, Kahler1991, Crooker1998, Shodhan2000}, it is reasonable to expect some degree of correlation between BDEs and magnetic complexity changes as well. 
To investigate this characteristic, we compare FR/ME categories and their complexity evolution history with information on the PAD of suprathermal electrons at 1~au (when available). 

In particular, we consider electron PAD data available at energies above $\sim100$~eV, i.e. in the suprathermal range.
For the events observed by ACE/SWEPAM, we use data in the 272~eV energy channel, while for events observed by the SWEA detector of STEREO/IMPACT, we integrate among the six energy channels operating above $\sim100$~eV, 
i.e. at 1716.84~eV, 1056.95~eV, 650.70~eV, 400.60~eV, 246.62~eV, and 151.83~eV. 
At STEREO, we compared the results obtained by integrating over all channels above $\sim100$~eV, with those obtained from the 246.62~eV channel alone (i.e. covering energies closest to those available at ACE), detecting no significant difference. We nevertheless considered it preferable to use multiple energy channels when possible, in order to ensure sufficient count rates for all periods under investigation.
For each event, PADs are normalized to the peak flux observed at each time step, in order to have distributions running between 0 and 1 throughout the ICME passage period.
The process of selecting BDE intervals within MEs is somewhat subjective, and becomes questionable for pitch angle distributions approaching isotropy and for cases where the flux of one of the counter-streaming beams is weaker than the other \citep{Shodhan2000}.
In this work, we apply the following identification procedure. 
For each ICME observed at 1~au, we determine the presence of BDEs within the ME or FR structure by identifying periods of counter-streaming particle beams located around $0^\circ$ and $180^\circ$ with respect to the nominal magnetic field direction. 
To account for cases where the two beams have different fluxes, we identify as BDEs periods of counter-streaming beams that are both reaching normalized fluxes of at least 0.5. As unbalanced suprathermal electron fluxes may indicate encounters with closed structures characterized by largely asymmetric legs, i.e. one of which extends out into the heliosphere well beyond the spacecraft location \citep{Pilipp1987}, placing a limit on the minimum flux of each beam may impose a limit to the maximum length of a closed magnetic field line before it is considered open \citep{Shodhan2000}. However, such an approach provides an unbiased, objective criterion to identify BDEs, hence for the purpose of this study, we consider the advantages to outweigh the limitations.

PAD data are scanned by eye, and periods approaching isotropy are discarded as non-BDE periods. We then classify BDE events according to the percentage of time during the FR (when present) or ME passage when counter-streaming electrons were observed. 
We bin the events into four categories: $75\%-100\%$ (BDEs throughout), $50\%-75\%$ (extended BDEs), $25\%-50\%$ (partial BDEs), and $0\%-25\%$ (no BDEs).
The presence or absence of BDEs for each ICME is provided in Table~S1.

\begin{figure*}
\centering
{\includegraphics[width=1\hsize]{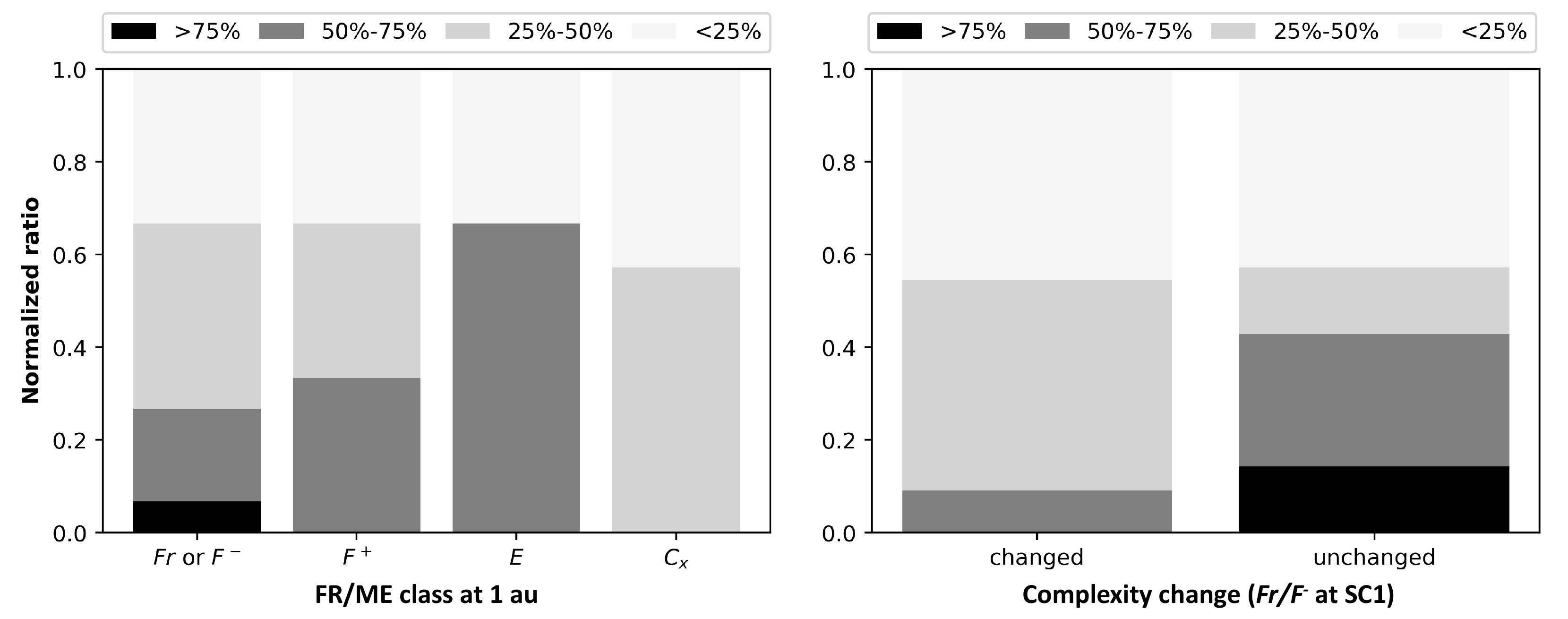}}
\caption{Occurrence of BDEs at 1~au as a function of the FR/ME category (left) and complexity evolution (right).} 
\label{fig:figure5}
\end{figure*}

In total, 28 ICMEs were observed at 1~au, all of which had associated suprathermal electron PAD data available.
We find that 64\% of the events at 1~au show some evidence of BDEs (partial to throughout), which is in good agreement with the 67\% estimate by \citet{Richardson2010}.
As shown in Figure~\ref{fig:figure5} (left), we find that 
33\% of $Fr/F^-$ types (out of a total of 15 events) exhibit no BDEs, 
40\% exhibit partial BDEs for less than half of the FR (when present) or ME duration, 
20\% exhibit extended BDEs, and 
7\% exhibit BDEs thoughout the whole structure. 
These percentages are only partially consistent with earlier (upper) estimates by \citet{Shodhan2000}, who found 
$17\%-27\%$ out of the 48 MCs considered (roughly corresponding to $Fr/F^-$ types in this study) did not contain BDEs for longer than 25\% of their duration, 
$10\%-21\%$ had partial BDEs, 
$15\%-27\%$ had extended BDEs,
and $29\%-42\%$ had BDEs throughout.
We argue that the discrepancies between our results and those reported by \citet{Shodhan2000} may be due to the different criteria used to identify BDEs, particularly in the case of unbalanced beam fluxes.
At the opposite end of the spectrum, we find that none among the 7 $C_x$ types at 1~au exhibited BDEs for more than half of the FR/ME structure, with 57\% of the cases exhibiting partial BDEs, and 43\% completely lacking BDEs. Finally, $F^+$and $E$ types were observed in a small number of cases (3 each), and thus no statistical conclusion can be drawn due to the low number of events available. The higher occurrence of prolonged BDEs within $Fr/F^-$ than within $C_x$ types is consistent with the interpretation of $Fr/F^-$ types typically reflecting closed, well-ordered FR structures \citep[e.g.][]{Zurbuchen2006}, while $C_x$ types reflect more complex magnetic topologies which likely generated through extensive magnetic interaction (implying, e.g., reconnection and reconfiguration of magnetic fields lines; \citet{Crooker1998, Winslow2016}) with their surroundings.

When considering the presence of BDEs with respect to the complexity evolution history of our events (Figure~\ref{fig:figure5}, right), we restrict ourselves to those events that were observed to have an $Fr/F^-$ configuration at the inner spacecraft, and neglect $F^+$, $E$, and $C_x$ types.
We consider only events whose initial magnetic structure is consistent with that of a closed, twisted FR rooted at the Sun as they are more likely to have had extended BDEs at the inner spacecraft according to the standard interpretation of BDEs within ICMEs \citep[e.g.][]{Zurbuchen2006}. 
This is done to ensure that the BDE properties observed at 1~au are actually the result of the evolution history between the inner and outer observing spacecraft, and not the result of evolution closer to the Sun.
We report 11 events that change complexity and 7 that do not change complexity. 
Of those that change complexity, only 1 (9\%) exhibit BDEs for more than a half of the FR/ME duration: $\sim 45$\% show partial BDEs, and $\sim 45$\% no BDEs whatsoever.
Among the events that do not change their complexity, 3 (43\%) exhibit throughout or extended BDEs, 1 (14\%) has partial BDEs, and 3 (43\%) have no BDEs. 
We note that of these 3 events (i.e. events 16, 18, 22), 2 exhibit prolonged unidirectional electron beams compatible with magnetic field lines open at one end and possibly originated by reconnection occurring close to the Sun \citep{Gosling1995, Crooker2002}, i.e. not by interplanetary sources. Only one event (event 18) among those that do not change complexity exhibit a highly-variable PAD including unidirectional and isotropic electron flows, potentially reflecting local and extended sources of reconnection.

Overall, prolonged BDEs are found to be more common among events that do not change complexity, i.e. that likely did not interact with other interplanetary structures. However, a similar fraction of events that change and that do not change complexity lack BDE signatures (43\% to 45\%). Interactions with the HCS seem to be the most disruptive condition for BDEs, consistent with a scenario of extensive magnetic reconnection taking place between the ICME and the interplanetary magnetic field near the HCS, as discussed for example by \citet{Winslow2016}. In conclusion, our results indicate that the duration of BDEs within FR/ME structures at the outer observing spacecraft does reflect the evolution history of ICMEs to some extent, but identifying complexity changes from these data products alone, even when in combination with magnetic field data, would be complicated by the large fraction of events lacking BDEs among both changed and unchanged populations. 
%

\subsubsection{ICME Speed and Magnetic Field at 1 au}
\label{subsubsec:speed_magnetic_field_1au}

We further investigate whether the complexity evolution history affects the internal properties at 1~au for a given ICME population. We focus on the relationship between the ICME speeds and magnetic fields at 1~au, which has been previously investigated because of its potential for the forecasting of geomagnetic storms \citep{Gonzalez1998, Owens2002, Owens2004, Owens2005}. 
Compared to previous studies, the novelty of our analysis lies in the investigation of ICME internal properties not only with respect to their magnetic field configuration, e.g. presence or absence of smooth magnetic field rotations within MEs, but also with respect to their complexity evolution and interaction histories.

\begin{figure*}
\centering
{\includegraphics[width=\hsize]{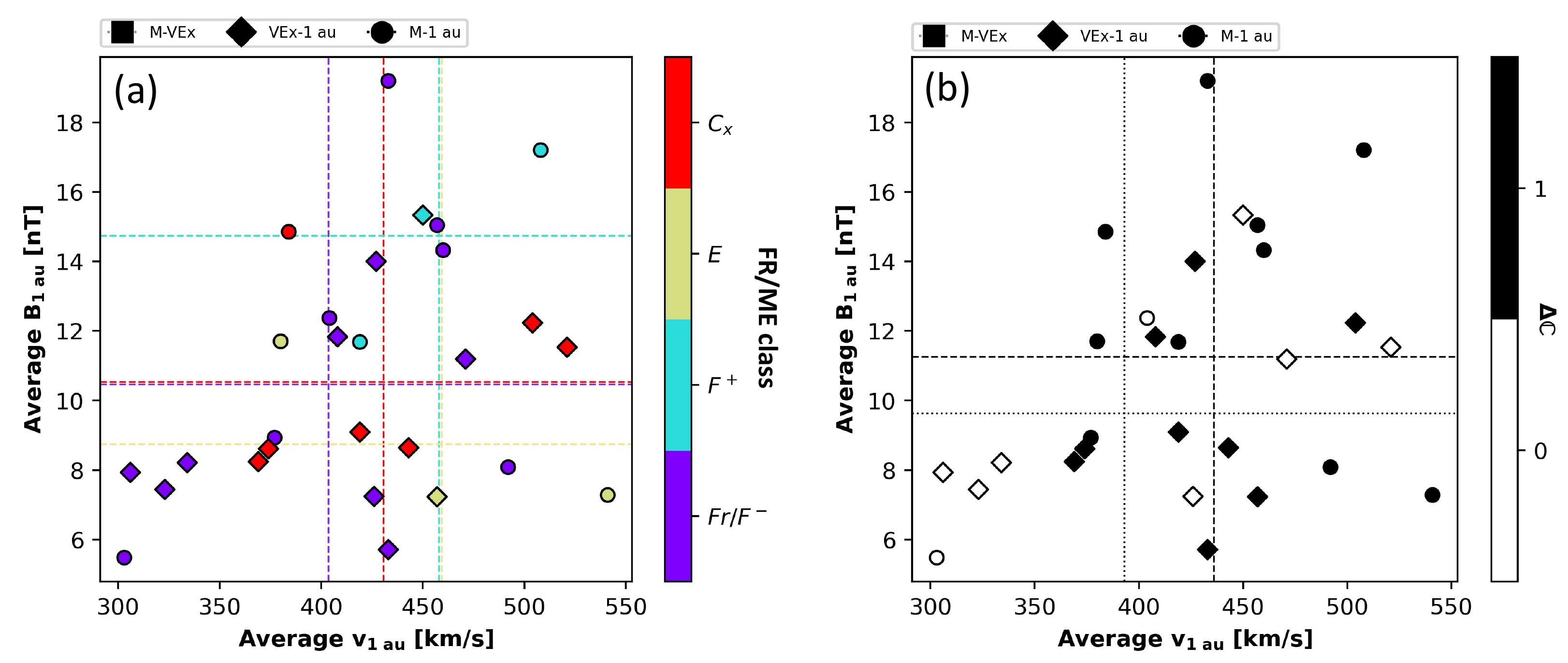}}
\caption{(a) FR/ME category and (b) complexity changes $\Delta \mathbb{C}$ as a function of the average speed and magnetic fields detected within ME and FR structures at 1~au. Different markers are used to indicate different pairs of spacecraft in radial alignment.}
\label{fig:figure6}
\end{figure*}

Figure~\ref{fig:figure6}~(a) shows the relationship between the mean speeds and magnetic fields observed at 1~au, by FR/ME category. 
$Fr/F^-$ and $C_x$ events (i.e. for which we have 15 and 7 events, respectively) have average speed values of 404~km~s$^{-1}$ and 431~km~s$^{-1}$, respectively. Yet, the application of the Welch’s t-test reveals no statistical significance in the mean speed of complex ($C_x$) and well-ordered ($Fr/F^-$) events, as indicated by the p-value of 0.36. 
The mean magnetic field observed at 1~au for both subsets is 10.5~nT.
We also report modest correlations between the ICME speed and magnetic field at 1~au for both $Fr/F^-$ and $C_x$ categories (Pearson's correlation coefficients, $PCC = 0.48$ and $0.28$, respectively).
These values are comparable to the correlation coefficients reported by \citet{Owens2005} for MC and non-MC ICMEs.

Figure~\ref{fig:figure6}~(b) shows the relationship between the mean speeds and magnetic fields observed at 1~au, this time categorized by the complexity evolution history of ICMEs. 
In this case, we have 9 events that do not change complexity, and 19 that do change complexity, and the average speed observed at 1~au is 393~km~s$^{-1}$ and 436~km~s$^{-1}$ for unchanged and changed events, respectively.
The mean magnetic field observed at 1~au is 9.6~nT and 11.3~nT for unchanged and changed events, respectively.
For these events, we note that similar trends in the mean magnetic field strengths are also observed at the inner observing spacecraft, where average values for unchanged and changed events are 36.4~nT and 60.1~nT at MESSENGER, and 15.3~nT and 19.4~nT at VEx, respectively.
However, the Welch’s t-test reveals there is no statistically significant difference in the mean values between the two groups at 1~au, as indicated by the p-values of 0.17 and 0.22 for the speed and magnetic field, respectively.
On the other hand, grouping events by their complexity evolution history reveals a high correlation between the ICME speed and magnetic field at 1~au for events that do not change their complexity ($PCC = 0.70$). The p-value is 0.04, indicating a 4\% probability for an uncorrelated system to produce datasets that have a $PCC$ at least as high as the one computed from the observed dataset of unchanged events. Further testing on the specific dataset considered reveals that by randomly picking 9 events among the 28 observed at 1~au (over 1 million realizations), there is an $\sim8$\% probability to produce datasets that have a $PCC$ at least as high as the one computed from the observed dataset of unchanged events. Overall, this indicates a $\sim92$\% confidence level that the speed and magnetic field within ICMEs that do not change their complexity are actually correlated. A correlation is not found for events that change their complexity ($PCC = 0.07$, with p-value of 0.79).

The high correlation between the ICME speed and magnetic field at 1~au observed for unchanged events, and the lack of correlation found for changed events, suggest that interaction with other solar wind structures (and, in turn, complexity changes) randomize the ICME internal properties, which, for unchanged events, appear relatively ordered. 
This may have important implications for space weather forecasting efforts, and for the usefulness of in situ magnetic field monitors located along the Sun--Earth line at inner heliocentric distances, as further discussed in Section~\ref{sec:conclusions}.

\subsubsection{ICME Magnetic Field Scaling with Heliocentric Distance}
\label{subsubsec:magnetic_field_scaling}

Another quantity that has been extensively investigated in previous studies is the scaling of the ICME magnetic field with heliocentric distance, which provides information on ICME global expansion \citep[e.g.][]{Demoulin2009, Lugaz2020b}.
Estimates of the decay of the average magnetic field inside MEs range between $-1.30 \pm 0.09$ and $-1.95 \pm 0.19$ \citep[e.g.][]{Liu2005, Wang2005, Leitner2007, Gulisano2010, Winslow2015, Davies2021b}, while estimates of the decay of the maximum magnetic field inside MEs range between $-1.73$ and $-1.89 \pm 0.14$ \citep[e.g.][]{Farrugia2005, Winslow2015}, with variations across different studied being the result of different ICME identification criteria and different ranges of heliocentric distances considered \citep{Gulisano2010}. All of the aforementioned studies considered statistical sets of ICMEs observed at different heliocentric distances, but not necessarily in radial alignment at multiple spacecraft.
\citet{Salman2020} did perform a fitting of the ME maximum magnetic field at different heliocentric distances for ICMEs observed at multiple spacecraft, finding a decay index of $-1.91$, but this value was calculated treating observations of the same ICME as independent from one another, and does not take into account specific trends underwent by individual events.
More recently, \citet{Lugaz2020b} estimated the average magnetic field decay within MEs by combining estimates obtained from individual ICMEs observed at multiple spacecraft in radial alignment. Using a set of 42 events taken from the catalog by \citet{Salman2020}, they reported decay indices of $-1.81 \pm 0.84$ and $-1.91 \pm 0.85$ for the maximum and average ME magnetic field, respectively. 
Considering a set of 18 FR ICMEs observed by radially-aligned spacecraft, \citet{Good2019} reported a scaling of the magnetic field along FR axes (as estimated by performing LFF fits to the observed FR magnetic structures) equal to $-1.34 \pm 0.71$.

\begin{figure*}
\centering
{\includegraphics[width=1\hsize]{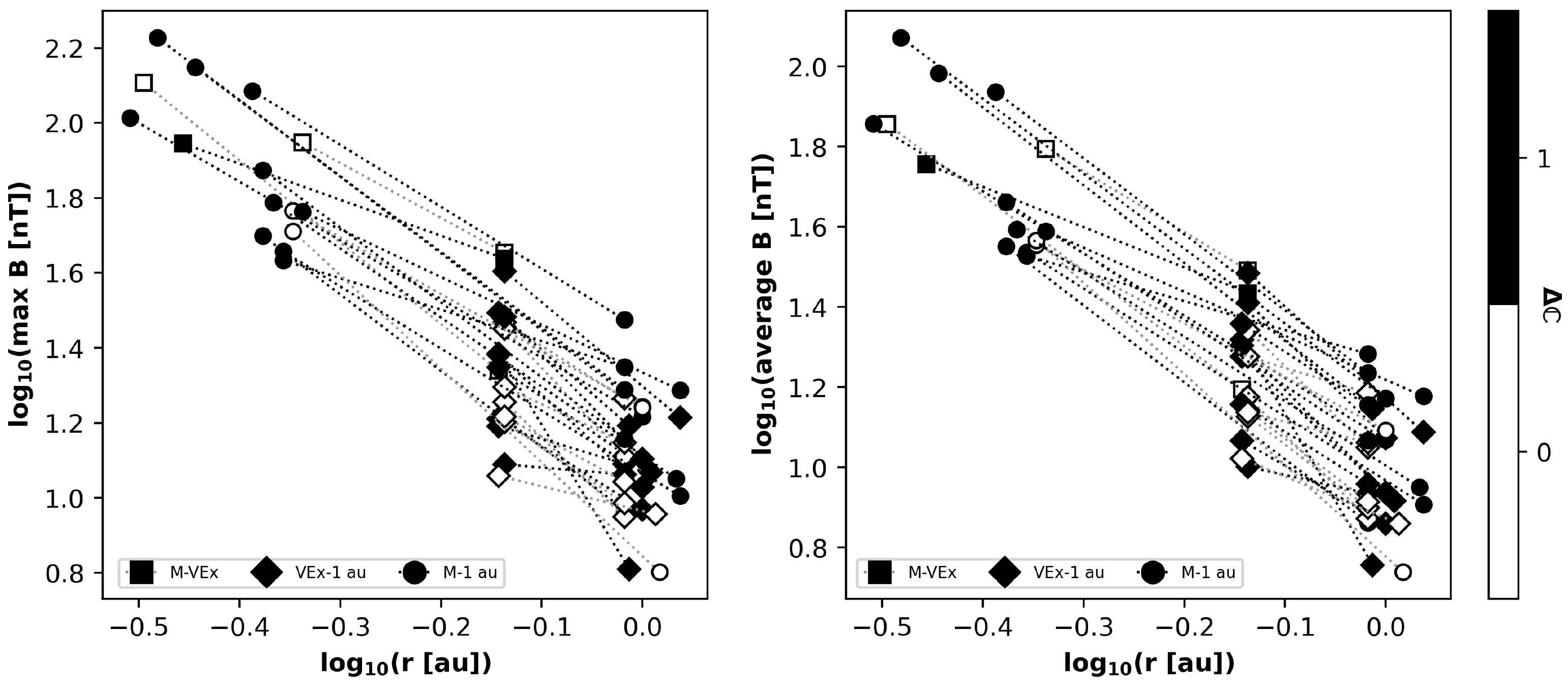}}
\caption{Complexity change $\Delta \mathbb{C}$ as a function of the radial evolution of the maximum (left) and mean (right) magnetic field within ME and FR structures. The points connected by a dotted line link observations of the same event at two spacecraft.
Different markers are used to indicate different pairs of spacecraft in radial alignment.}
\label{fig:figure7}
\end{figure*}

Similarly to \citet{Good2019} and \citet{Lugaz2020}, we estimate the decay index of the average and maximum magnetic field within MEs with heliocentric distance by taking the average of the decay indices computed for individual events (Figure~\ref{fig:figure7}). We provide the standard deviation as a measure of the uncertainty.
We find that the mean magnetic field scales as $-1.81 \pm 0.78$ and $-1.66 \pm  0.63$ for events that change and do not change their complexity, respectively. 
The maximum magnetic field scales as $-1.84 \pm 0.85$ and $-1.82 \pm 0.62$ for events that change and do not change their complexity, respectively. 
Overall, these numbers are consistent with previous studies and do not highlight any significant difference in the typical radial scaling of the magnetic field for events with different evolution histories.
We further apply the Welch’s t-test to determine if differences between events that change and that do not change complexity during propagation are statistically significant. We find p-values of 0.62 and 0.92 for the mean and maximum magnetic field slopes, indicating no statistically significant difference in the mean values between the two groups.
Yet, we note that the decay indices of events that change their complexity during propagation are characterized by larger standard deviations, indicating distributions more broadly-scattered around the mean compared to those of unchanged events. This result is consistent with the interpretation of a randomization of the ICME properties induced by interactions with other solar wind structures, as discussed in Section~\ref{subsubsec:speed_magnetic_field_1au}.

\subsubsection{FR-to-ME Duration Scaling with Heliocentric Distance}
\label{subsubsec:duration_scaling}

Finally, we investigate the relationship between ME and FR boundaries in terms of duration and its evolution with heliocentric distance. 
To the best of our knowledge, this aspect of ICME evolution has never been investigated before, and it may provide valuable information on magnetic erosion, occurring when magnetic reconnection between FR-contained plasma and the surrounding interplanetary magnetic field ``peels away'' the FR outer layers \citep{Dasso2006, Ruffenach2012}. Magnetic erosion affects $\sim$30\% of the ICMEs at 1~au, and it can be most reliably identified from the presence of reconnection exhausts and bifurcated current sheets near FR boundaries \citep{Ruffenach2015}. Erosion also implies large-scale magnetic field topological changes occurring within ICMEs \citep{Ruffenach2012, Pal2021}, and yet, the full breadth of effects on magnetic field configurations within MEs and FRs therein remains unclear.
In the following, we investigate the relationship between ME and FR duration as a potential proxy for magnetic erosion.

\begin{figure*}[t!]
\centering
{\includegraphics[width=0.55\hsize]{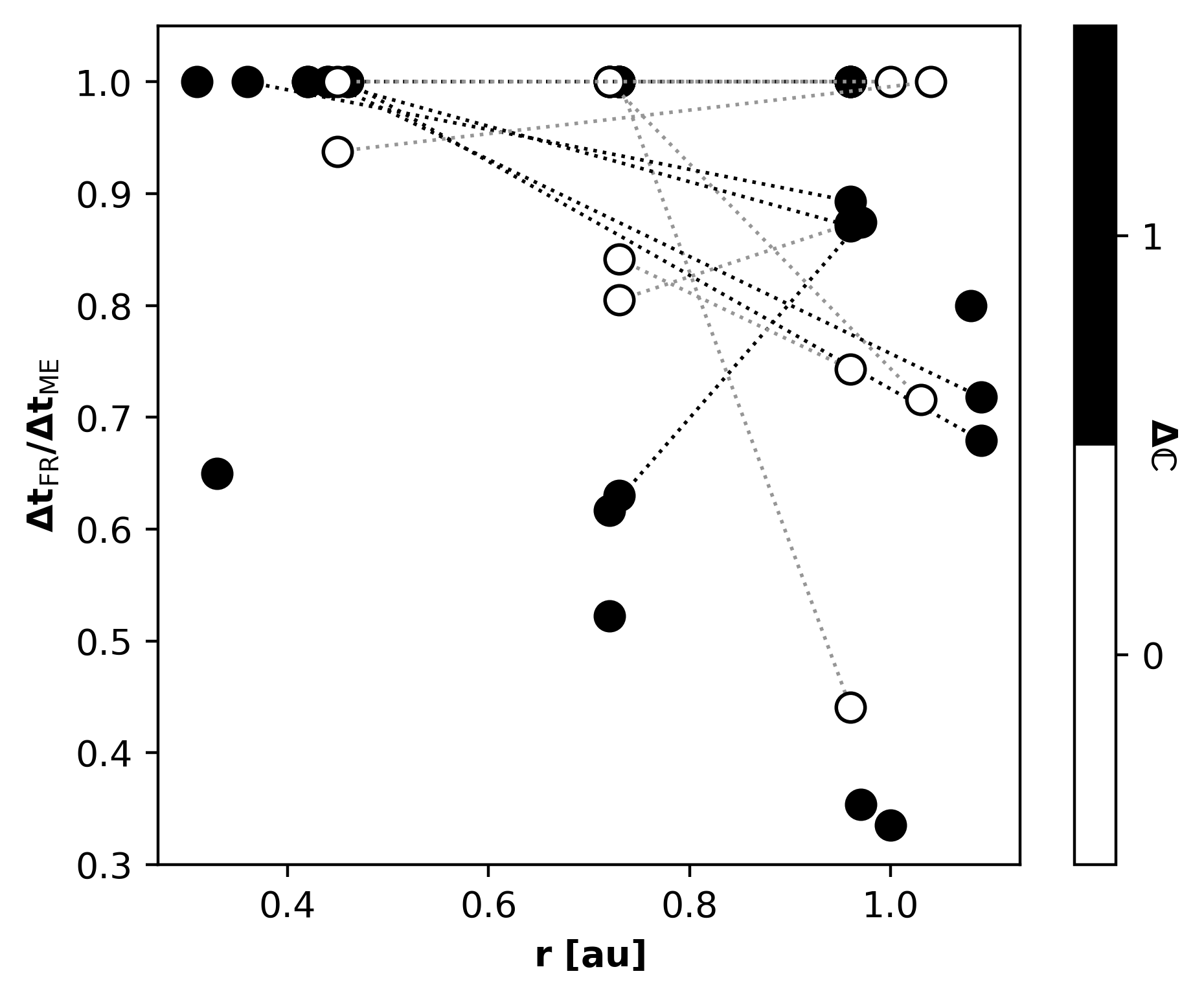}}
\caption{Complexity change $\Delta \mathbb{C}$ as a function of the radial evolution of the ratio between the FR duration ($\Delta t_\mathrm{FR}$) and ME duration ($\Delta t_\mathrm{ME}$) for the events exhibiting $F$-type magnetic structures at a given observing spacecraft. The dotted lines connect observations of the same event at two spacecraft, while points not connected by lines indicate ICMEs observed as $F$ types only at one observing spacecraft.} 
\label{fig:figure8}
\end{figure*}

Figure~\ref{fig:figure8} shows the radial evolution of the ratio between the FR duration ($\Delta t_\mathrm{FR}$) and ME duration ($\Delta t_\mathrm{ME}$) for the events exhibiting an FR magnetic structure (i.e. $Fr$, $F^-$, $F^+$ configurations) at least at one observing spacecraft. 
Overall, 27 ICMEs exhibit an FR structure at least at one observing spacecraft.
We find that 18 of them ($\sim58\%$ of the 31 ICMEs considered in total) exhibit an FR structure that is shorter than the whole ME structure at least at one observing spacecraft. Among the 62 detected ICME signatures (i.e. two per ICME), 22 ($\sim25\%$) have this characteristic. At 1~au, the percentage of signatures exhibiting an FR-to-ME duration ratio below 1 is $\sim46\%$. 
On average, the ratio between the FR and ME duration decreases with heliocentric distance from 0.96 at MESSENGER (number of observations: 11), to 0.88 at VEx (number of observations: 13), and 0.79 at 1~au (number of observations: 18), indicating FRs tend to shorten with respect to MEs during propagation. This decreasing trend is also reflected in the minimum values recorded at each heliocentric distance, which decreases from 0.65 at MESSENGER, to 0.52 at VEx and to 0.32 at 1~au. 
We note that while identifying ME and FR boundaries as described in Section~\ref{sec:methods}, we tried to treat the magnetic field data at 1~au in an as consistent manner as possible to the magnetic field data at inner distances, in order to minimize the bias due to the additional availability of plasma information at 1~au compared to MESSENGER and VEx data.
The behavior of individual events varies, but except for one event (number 25), all ICMEs are found to either decrease or maintain this ratio approximately constant (within $\pm 10\%$), regardless of their complexity evolution history.
Moreover, we note that events that change their complexity are consistently able to reach shorter minimum FR-to-ME duration ratios than events that do not change their complexity, at all heliocentric distances considered.
Overall, our findings indicate that FR shortening is ubiquitous and it affects both ICMEs that do change as well as those that do not change their magnetic complexity during propagation.

%
The investigation of the relationship between the observed FR-to-ME shortening, magnetic erosion, and suprathermal electron PADs discussed in Section~\ref{subsubsec:results_BDEs} goes beyond the scope of this work, and is left for future studies. Nevertheless, the fact that FRs shorten compared to MEs during propagation even for events that do not change their complexity may indicate that magnetic reconnection at the FR edges is not necessarily inducing complexity changes, and may naturally happen during propagation irrespective of the complexity evolution and interaction history of a particular ICME.

We also acknowledge two major limitations for statistical investigations of FR-to-ME shrinking and magnetic erosion phenomena from multi-spacecraft ICME observations based on existing data at different heliocentric distances. 
First, the lack of information on the plasma conditions (e.g. bulk properties as well as particle PADs), together with the numerous data gaps and bow shock and magnetospheric crossings affecting MAG data at MESSENGER and VEx, complicate the identification of ME front and rear boundaries at inner heliocentric distances \citep[see e.g.][for a detailed analysis of an ICME sheath/ME boundary at MESSENGER]{Lugaz2020}. Second, high-cadence plasma data is also critical to identify signatures of magnetic reconnection (e.g. the presence of current sheets and Alfv\'{e}nic fluctuations), which is likely the prime physical phenomenon responsible for the observed FR shortening. Such data is also currently missing for most ICME events within 1~au.
At the time of writing, the arrival of new data from e.g., Parker Solar Probe \citep{Fox2016}, Solar Orbiter \citep{Mueller2020} and BepiColombo \citep{Benkhoff2010} is already providing essential additions to pre-existing datasets \citep{Moestl2021}. 
Yet, current solar minimum conditions may require a number of years before a sufficient number of ICME magnetic structures are observed in radial alignment with a statistically-significant coverage of a broad spectrum of heliocentric distance combinations \citep{Moestl2020}.

\section{Summary and Conclusions}
\label{sec:conclusions}

In this paper, we performed the first statistical analysis of magnetic complexity changes occurring within ICMEs observed by multiple inner heliospheric spacecraft in radial alignment.
Our aim was to answer the following main open questions: 
(1) How frequently do ICMEs undergo magnetic complexity changes during propagation through interplanetary space? 
(2) What are the causes of such changes?
(3) Do the in situ properties of ICMEs differ depending on whether they exhibit complexity changes during propagation?  
To address these questions, we considered multi-spacecraft observations of 31 ICMEs from MESSENGER, VEx, ACE, and the STEREO spacecraft between December 2008 and February 2014 during periods of radial alignment.
By analyzing the changes in the ICME magnetic field properties between the inner and outer observing spacecraft, we were able to identify magnetic complexity changes which manifested as fundamental alterations of the magnetic configuration, or as significant re-orientations of the structure. From the inspection of plasma and suprathermal electron PAD data at 1~au, and simulations of the ambient solar wind performed with the WSA-ENLIL model, we reconstructed the propagation scenario for each of the 31 events, and identified the decisive factors determining their evolution between the inner and outer observing spacecraft.
The main results can be summarized as follows:

\begin{enumerate}

\item We found that, on average, ICMEs tend to increase their magnetic complexity with radial distance. MEs with $Fr/F^-$ configurations are less frequently observed at 1~au ($\sim53\%$) than they are at Mercury ($\sim67\%$), while complex $C_x$ types appear more frequently at 1~au ($\sim25\%$) than at Mercury ($\sim13\%$). 
We stress that the detection of an FR configuration at 1~au is not sufficient to conclude that a given ICME did not undergo any complexity change earlier during propagation. In fact, many FRs at 1~au are found to have gone through significant changes at closer heliocentric distances.
Overall, results indicated that roughly $42\%$ of the events undergo drastic alterations of their magnetic configuration during propagation, and $63$\% exhibit significant changes in their magnetic complexity, including major rotations of FR structures. Magnetic complexity changes within ICMEs therefore occur more frequently than previously estimated, and they in fact affect the majority of ICMEs between Mercury and 1~au. 
\item Based on our results, ICMEs tend to preserve their magnetic topology and orientation only over short spatial and temporal scales, i.e. when observed in close radial alignment ($\Delta \phi <15^\circ$) and at radial separations $\Delta r \leq 0.4$~au. At such separations, the fraction of ICMEs preserving their complexity is $\sim 90\%$. This fraction drops to $\sim 29\%$ for observations at $\Delta r>0.4$~au. Similar fractions are also observed by spacecraft in perfect radial alignment ($\Delta \phi < 5^\circ$). These results suggests that propagation (``nurture'') effects are likely the main cause of complexity changes observed by spacecraft separated by $\Delta \phi <15^\circ$.
Observations at $\Delta \phi > 15^\circ$ detected complexity changes for all events, regardless of the spacecraft radial separation. In principle, this result could be interpreted as a combination of different factors: 
pre-existing inhomogeneities in the internal structure of ICMEs (``nature'' effects), 
variable ambient solar wind conditions through which different parts of ICMEs propagate, 
and ultimately, changes during propagation due to interaction with other structures (both ``nurture'' effects). 

However, the majority of complexity changes involving significant re-orientations of ICME FRs were detected by spacecraft separated by less than $15^\circ$ in longitude, indicating that magnetic complexity changes identified through the application of Condition~B (previously defined in Section~\ref{subsubsec:complexity_change_def_B}) cannot be explained in terms of large-scale geometric effects due to spacecraft crossings through different parts of stretched FRs.
In most of the cases considered, we also found that fundamental alterations to FR/ME structures as defined by Condition~A (previously defined in Section~\ref{subsubsec:complexity_change_def_A}) were associated with different kinds of interactions, suggesting the relatively large angular separations among the observing spacecraft was not the primary cause of such changes. 
Overall, these arguments suggest that for the majority of the events considered, the results reflect actual magnetic complexity changes affecting ICMEs, rather than geometrical effects due to the particular spacecraft trajectories along a given structure.
Additionally, despite the caveat of non-negligible radial separations affecting our observations, we found a $\Delta \phi \sim 15^\circ$ breaking point which likely provides a lower limit to the characteristic scale of ICME magnetic coherence, and it supports previous estimates set to $17^\circ - 26^\circ$ and $14^\circ - 20^\circ$ by \citet{Owens2020} and \citet{Lugaz2018} based on theoretical and observational arguments, respectively.
Reconstructions of the inclination of magnetic flux rope structures using a LFF fitting technique over a subset of events considered in this study also indicate that the aforementioned results apply to low-inclination flux ropes. Establishing possible effects of inclination on ICME magnetic complexity changes and estimates of ICME magnetic coherence will require the consideration of highly-inclined ICME magnetic flux ropes by future studies.
\item We found that ICME magnetic complexity changes are tightly related to their interaction with solar wind structures in interplanetary space. Although such kind of interactions are not always inducing a detectable change in ICME complexity, they do increase the probability for ICMEs to undergo a complexity change. On the one hand, most ICMEs that preserved their magnetic configuration during propagation ($\sim90\%$) either did not interact with any interplanetary structure, or only interacted with a single structure. On the other hand, events that changed their magnetic complexity were more likely to have interacted with two or more structures (in $\sim 45\%$ of the cases considered). We also report that $C_x$-class, complex MEs are always associated with ICME--solar wind interactions, particularly with the HCS (in $\sim 57\%$ of the cases). 
%
%
\item At 1~au, we were able to identify prolonged BDEs within $\sim 25\%$ of the MEs characterized by a twisted flux-rope configuration, while prolonged BDEs are completely absent within $C_x$ types. This supports the interpretation of $Fr/F^-$ types often being characterised by closed magnetic loops rooted at the Sun, while $C_x$ types may be fundamentally different structures likely generated through extensive magnetic reconfiguration with the surroundings.
Prolonged BDEs are also more commonly found within events that do not change complexity during propagation (in 43\% of the cases compared to 9\% of changed events).
Our results suggest that the duration of BDE flows detected from suprathermal electron PAD data at the outer observing spacecraft do reflect the evolution history of ICMEs, but considered alone, they would not be sufficient to identify magnetic complexity changes.
Additionally, the low percentage of prolonged BDEs even within the events showing no complexity change indicates there might be in fact even more complexity changes underwent at closer heliocentric distances, which we were unable to quantify due to the lack of PAD data closer to the Sun.
\item We reported a significant correlation ($PCC=0.70$) between the mean speed and magnetic field detected at 1~au for ICMEs that preserved their magnetic complexity during propagation. This correlation is lost in the case of events that undergo complexity changes ($PCC=0.07$), indicating the interaction with solar wind structures likely randomizes the average ICME internal properties, making them hardly predictable. 

The correlation observed for unchanged events revives previous attempts to apply speed-magnetic field correlations to space weather forecasting \citep{Owens2005}, and it may have important implications for the usefulness of in situ magnetic field monitors located along the Sun--Earth line at inner heliocentric distances.
The achievement of more accurate and timely predictions of the arrival time and location of SIRs and the HCS by solar wind models, particularly in the Earth-facing hemisphere, is being boosted by the launch of new heliospheric missions, the availability of increased numerical capabilities, and the application of novel techniques such as data assimilation, machine learning, and ensemble modeling \citep[e.g.][]{Gressl2014, Reiss2019, Reiss2020, Szabo2020, Bailey2021, Owens2021, Samara2021}. Such methods can allow prompt identification of ICMEs that are going to propagate through a quiet solar wind environment, with computationally-fast ICME propagation models such as drag-based models that can further provide reliable predictions of the ejecta speed at 1~au \citep{Calogovic2021}. Our results suggest that such a combination of tools may therefore be sufficient to provide a reliable estimate of the ejecta magnetic field at 1~au for ICMEs propagating through a quiet environment. 
Moreover, by our definition (Section~\ref{subsec:complexity_changes}), unchanged events do not undergo significant restructuring nor rotation of their internal magnetic structure during propagation through interplanetary space. In the near future, remote-sensing observations of CME magnetic fields during early propagation phases \citep[such as coronal magnetic field measurements provided by DKIST;][]{Rast2021} may allow accurate extrapolations of the ICME magnetic field orientation to 1~au, at least for those events expected not to interact with any structure during propagation to a given impact location.
If used in combination, these two approaches may bring significant improvement to prediction capabilities not only of the ICME magnetic field magnitude at the impact location, but also of its orientation (including the most geo-effective $B_z$ magnetic field component).
Similarly, early measurements of the ICME magnetic field orientation performed in situ by spacecraft orbiting upstream of Earth as close as 0.3~au from the Sun may provide accurate predictions of their magnetic field direction at Earth. 
In turn, our results set a caveat on the use of inner heliospheric observations upstream of the Earth to predict the geoeffectiveness of ICMEs in the presence of large interplanetary structures in the ICME propagation space.
The scenario is more complicated at times when solar wind structures are present in the ICME transit path from the Sun to 1~au. 
In such cases, neither estimates of the speed at 1~au, nor situ magnetic field data upstream of the Earth would be sufficient to accurately estimate the magnetic field strength and orientation at the impact location.
Our results provide a statistical confirmation to earlier suggestions by \citet{Winslow2016} based on a single ICME case study. 
\item On average, ICMEs with different evolution histories are found to behave similarly when considering the scaling of the FR/ME magnetic field with heliocentric distance. However, we reported a wider variety of magnetic field radial behaviors among events that changed their magnetic complexity, supporting the interpretation that interactions with other interplanetary structures drive major changes to the internal ICME properties, randomizing their characteristics and complicating their prediction at given target locations. Additionally, FR structures most often shorten with respect to MEs during propagation from Mercury to 1~au, regardless of whether a particular ICME changes or does not change its magnetic configuration. 
We interpret this result as an indication that magnetic reconnection, and particularly magnetic erosion at the front and rear of MEs, does not necessarily induces complexity changes. In fact, FR erosion likely naturally happen during propagation irrespective of the complexity evolution and interaction history of a particular ICME.
\end{enumerate}

The consideration of a statistical set of ICMEs allowed us to generalize previous results by \citet{Winslow2016, Winslow2021} and to draw, for the first time, conclusions on the frequency, causes, and effects of magnetic complexity changes on ICMEs.
Most importantly, such an investigation provided evidence of the ubiquitousness of complexity changes affecting ICMEs propagating throughout the inner heliosphere, and allowed us to identify interactions with other large-scale solar wind structures as their primary drivers. 
This result places magnetic complexity changes as a consequence of ICME interactions with large-scale interplanetary structures in their surroundings, rather than as intrinsic to ICME evolution during propagation.

%
The details of how magnetic complexity changes propagate throughout ICME magnetic structures, as well as of the physical mechanisms mediating such changes, remain open questions to be addressed in future studies.
In this respect, despite the observational limitations affecting past missions (e.g. the uneven and inhomogenous datasets available at different heliocentric distances, and the low number of radially-aligned observations of rare ME configurations, such as $F^+$ and $E$ categories), new observational sources will help unlock additional areas of investigation in the future. 
%
Over the short and medium term, the availability of both high-cadence magnetic field and plasma data from recently-launched heliospheric and planetary missions, such as Parker Solar Probe, Solar Orbiter, and BepiColombo, will open up new windows of opportunity for dedicated studies on the effect of interactions on ICME magnetic structures and their large-scale topology and connectivity, especially for selected case studies that will be observed within the current solar cycle. 
In the longer-term perspective of achieving more comprehensive and statistically-valid results regarding the evolution of ICMEs throughout the inner heliosphere, e.g. throughout the exploration of the full parameter space of spacecraft radial distances and separations, we also stress the importance of initiatives aimed at increasing the number of missions equipped with magnetometers and plasma instruments orbiting at different distances from the Sun. 


\acknowledgments 
Support for this work was provided by NASA grant 80NSSC19K0914. 
C.S.\ acknowledges the NASA Living With a Star Jack Eddy Postdoctoral Fellowship Program, administered by UCAR's Cooperative Programs for the Advancement of Earth System Science (CPAESS) under award no.\ NNX16AK22G.\  
R.M.W. and E.E.D acknowledge support from NASA grant 80NSSC19K0914, and R.M.W. also acknowledges partial support from the NASA STEREO TRANSITION grant. 
N.L. acknowledges support from NASA grants 80NSSC20K0431 and 80NSSC20K0700.

All the data analyzed in this study are publicly available. 
ACE data have been obtained from the Space Physics Data Facility’s Coordinated Data Analysis Web (\url{https://cdaweb.gsfc.nasa.gov/}) and the ACE Science Center (\url{http://www.srl.caltech.edu/ACE/ASC/}).
STEREO data have been downloaded from the STEREO Data Server at UCLA (\url{https://stereo-dev.epss.ucla.edu/l1_data/}), the Space Physics Data Facility’s Coordinated Data Analysis Web (\url{https://cdaweb.gsfc.nasa.gov/}), and the IRAP SWEA server (\url{http://stereo.irap.omp.eu/CEF/PAD/}).
MESSENGER data have been obtained from the Space Physics Data Facility’s Coordinated Data Analysis Web (\url{https://cdaweb.gsfc.nasa.gov/}), while Venus Express data are available on the Planetary Science Archive (PSA) at ESA (\url{https://archives.esac.esa.int/psa/}).

%






\bibliographystyle{aasjournal}



\end{document}